\documentclass[preprint,authoryear,12pt]{elsarticle}

\usepackage{geometry} % see geometry.pdf on how to lay out the page. There's lots.
\usepackage{graphicx}
\usepackage{subfigure}
\usepackage{booktabs}
\usepackage{color,amssymb}
\usepackage{setspace}
\usepackage{appendix}
\usepackage{longtable, lscape}
\usepackage{latexsym,amssymb,enumerate,amsmath,bm}

\newcommand{\beq}{\begin{equation}}
\newcommand{\eeq}{\end{equation}}

\newcommand{\vs}{\left( \begin{array}{c}}
\newcommand{\ve}{\end{array}\right)}

\newcommand{\ms}{\left( \begin{array}}
\newcommand{\me}{\end{array}\right)}

\biboptions{comma,square, sort}
\journal{Social Networks}

\begin{document}

\title{Social Structure of Facebook Networks}

\author{Amanda L. Traud$^{1,2}$, Peter J. Mucha$^{1,3}$, and Mason A. Porter$^{4,5}$\\
  $^1$\footnotesize{Carolina Center for Interdisciplinary Applied Mathematics, Department of Mathematics,\\ University of North Carolina, Chapel Hill, NC 27599-3250, USA} \\
  $^2$\footnotesize{Carolina Population Center, University of North Carolina, Chapel Hill, NC 27516-2524, USA} \\
  $^3$\footnotesize{Institute for Advanced Materials, Nanoscience \& Technology,\\
  University of North Carolina, Chapel Hill, NC 27599-3216, USA} \\
  $^4$\footnotesize{Oxford Centre for Industrial and Applied Mathematics, Mathematical Institute,\\ University of Oxford, OX1 3LB, UK}\\
  $^5$\footnotesize{CABDyN Complexity Centre, University of Oxford, OX1 1HB, UK}\\
}

\maketitle

%%%%%%%%%%

\section*{Abstract}

We study the social structure of Facebook ``friendship''  networks at one hundred American colleges and universities at a single point in time, and we examine the roles of user attributes---gender, class year, major, high school, and residence---at these institutions.  We investigate the influence of common attributes at the dyad level in terms of assortativity coefficients and regression models.  We then examine larger-scale groupings by detecting communities algorithmically and comparing them to network partitions based on the user characteristics. We thereby compare the relative importances of different characteristics at different institutions, finding for example that common high school is more important to the social organization of large institutions and that the importance of common major varies significantly between institutions.  Our calculations illustrate how microscopic and macroscopic perspectives give complementary insights on the social organization at universities and suggest future studies to investigate such phenomena further.

%%%%%%%%%%%

\section{Introduction}

Since their introduction, social networking sites (SNSs) such as Friendster, MySpace, Facebook, Orkut, LinkedIn, and myriad others have attracted hundreds of millions of users, many of whom have integrated SNSs into their daily lives to communicate with friends, send e-mails, solicit opinions or votes, organize events, spread ideas, find jobs, and more \citep{boydell}.  Facebook, an SNS launched in February 2004, now overwhelms numerous aspects of everyday life, and it has become an immensely popular societal obsession 
%among college and high school students (and, increasingly, among other members of society) 
\citep{boydheart,boydell,lewis,oldboy}.  Facebook members can create self-descriptive profiles that include links to the profiles of their ``friends," who may or may not be offline friends.  Facebook requires that anybody who one wants to add as a friend confirm the relationship, so Facebook friendships define a network (graph) of reciprocated ties (undirected edges) that connect individual users.

The emergence of SNSs such as Facebook and MySpace has revolutionized the availability of social and demographic data, which has in turn had a significant impact on the study of social networks \citep{boydell,orgnet,newmedia}.  It is possible to acquire very large data sets from SNSs, though of course the population online and actively using SNSs is a biased sample of the broader population.  Services like Facebook also contain large quantities of demographic data, as many users now voluntarily reveal voluminous amounts of detailed personal information.  An especially exciting aspect of studying SNSs is that they provide an opportunity to examine social organization at unprecedented levels of size and detail, and they also provide new venues to test sampling effects \citep{kurant2011}. One can investigate the structure of an SNS like Facebook to examine it as a network in its own right, and ideally one can also try to take one step further and infer interesting insights regarding the offline social networks that an SNS imperfectly parallels.  Most people tend to draw their Facebook friends from their real-life social networks \citep{boydell}, so it is not entirely unreasonable to use Facebook networks as a 
%imperfect
proxy for an offline social network.  (Of course, as noted by \cite{hogan09}, one does need to be aware of significant limitations when taking such a leap of faith.)

Social scientists, information scientists, and physical scientists have all jumped on the SNS data bandwagon \citep{nyt}. It would be impossible to exhaustively cite all of the research in this area, so we only highlight a few results; additional references can be found in the review by \cite{boydell}.  \cite{boyd} also wrote a popular essay about her empirical study of Facebook and MySpace, concluding that Facebook tends to appeal to a more elite and educated cross section than MySpace.  The company RapLeaf \citep{rapleaf} has compiled global demographics on the age and gender usage of numerous SNSs.  Other recent studies have investigated the manifestation on SNSs of race and ethnicity \citep{gajjala}, religion \citep{nyland}, gender \citep{geidner,hjorth}, and national identity \citep{fragoso}.  Preliminary research has also suggested that online friendship networks can be exploited to improve shopper recommendation systems on websites such as Amazon
\citep{filter}.

Several papers have attempted to increase understanding of how SNS friendships form.  For example, \cite{kumar} examined preferential attachment models of SNS growth, concluding that it is important to consider different classes of users. \cite{lampe} explored the relationship between profile elements and number of Facebook friends, and other scholars have examined the importance of geography \citep{liben} and online message activity \citep{golder} to online friendship formation.  Other papers have established strong correlations between network participation and website activity, including the motivation of people to join particular groups \citep{backstrom}, the recommendations of online groups \citep{spertus}, online messages and friendship formation \citep{golder}, interaction activity versus sense of belonging \citep{alvin07}, and the role of explicit ideological relationship designations in affecting voting behavior \citep{szabo1,szabo2}.   \cite{lewis} used Facebook data for an entire class of freshmen at an unnamed, private American university to conduct a quantitative study of social networks and cultural preferences. The same data set was also used to examine user privacy settings on Facebook \citep{lewis2}. 

In the present paper, we study the complete Facebook networks of 100 American college and universities from a single-day snapshot in September 2005.  This paper is a sequel to our previous research on 5 of these institutions \citep{siamface}, in which we developed some of the methodology that we employ here.  In September 2005, one needed a {\tt .edu} e-mail address to become a member of Facebook, and the majority of ``friendship'' ties were within the same institution.  We thus ignore links between nodes at different institutions and study the Facebook networks of the 100 institutions as 100 separate networks.  For each network, we have categorical data encompassing the gender, major, class year, high school, and residence (e.g., dormitory, House, fraternity, etc.) of the users.  
We examine homophily and community structure (network partitions that are obtained algorithmically) for each of the networks and compare the community structure to partitions based on the given categorical data. We thereby compare and contrast the organizations of the 100 different Facebook networks, which arguably allows us to compare and contrast the organizations of the underlying university social networks that they imperfectly represent.  In addition to the inherent interest of these Facebook networks, our investigation is important for subsequent use of these networks---which were formed via ostensibly the same generative mechanism online---as benchmark examples for numerous types of computations, such as new community detection methods.

The remainder of this paper is organized as follows.  We first discuss the Facebook data and present the methods that we used for testing homophily at the dyad level and demographic prevalences at the community level. We then present and discuss results on the largest connected components of the networks, student-only subnetworks, and single-gender subnetworks.  Finally, we summarize and discuss our findings.

%%%%%%%%%%%%%

\section{Data}

The data that we use was sent directly to us in anonymized form by Adam D'Angelo of Facebook.  It consists of the complete set of users (nodes) from the Facebook networks at each of 100 American institutions (which we enumerate in Table \ref{tab:CharTable}) and all of the ``friendship'' links between those users' pages as they existed in September 2005. The data clearly identifies most institutions, although there are a small number of disambiguation problems.  For instance, 4 different ``UC'' institutions plus ``Cal'' are in the data, and there are 2 ``Texas'' listings. Each institution in the data includes a number appearing as part of its name that appears to correspond to the order in which each institution ``joined'' Facebook. The data can be downloaded at {\tt http://people.maths.ox.ac.uk/$\sim$porterm/data/facebook100.zip}.

Similar snapshots of Facebook data from 10 Texas institutions were analyzed recently by \cite{oldboy}, and a snapshot from ``a diverse private college in the Northeast U.S.'' was studied by \cite{lewis}.  Other studies of Facebook have typically obtained data either through surveys \citep{boydell} or through various forms of automated sampling \citep{ucifacebook}, thereby missing nodes and links that can impact the resulting graph structures and analyses.  We consider only ties between people at the same institution, yielding 100 separate realizations of university social networks and allowing us to compare the structures at different institutions.

\begin{figure}[htbp]
\centerline{
\includegraphics[width=1.0\textwidth]{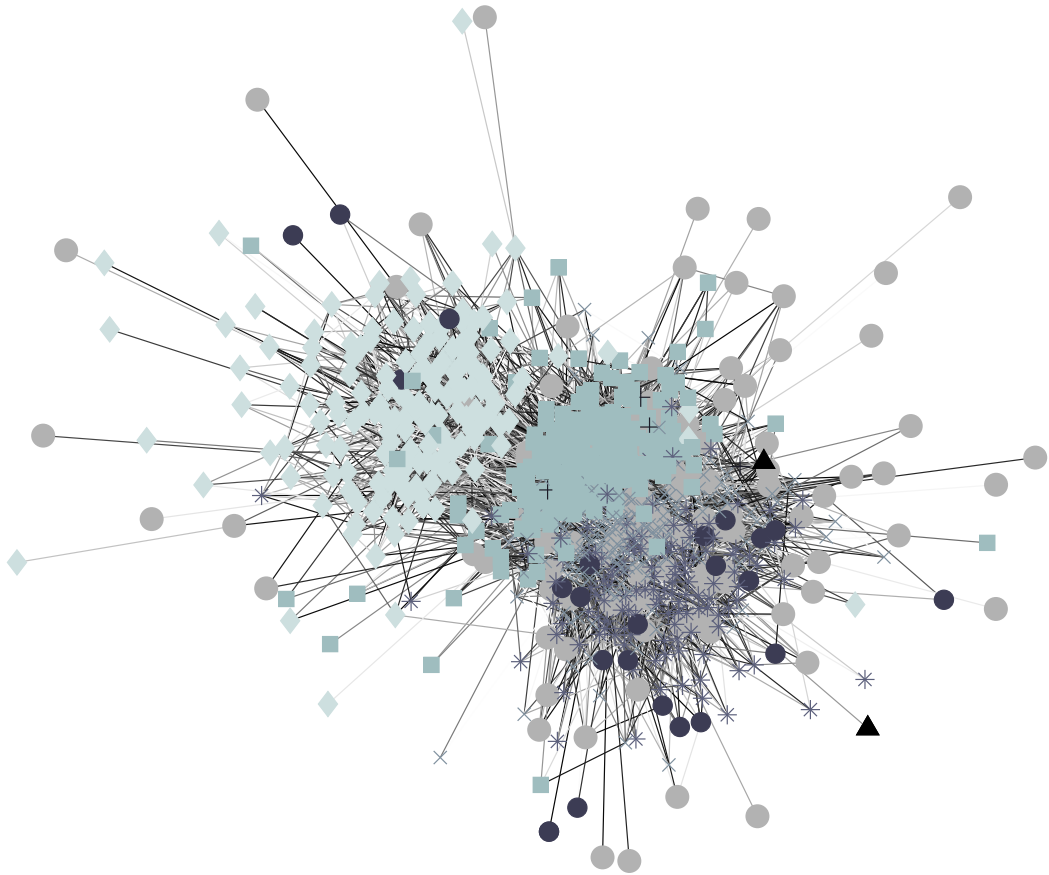}
		}
\caption{Largest connected component of the student-only subset of the Reed College Facebook network. (We used a \cite{FruchtermanReingold91} visualization.) Different node shapes and gray scale indicate different class years (gray circles denote users who did not identify an affiliation), and the edges are randomly shaded for easy viewing. Clusters of nodes with the same grayscale/shape suggest that common class year has an important effect on the aggregate Facebook structure. 
}
\label{viz}
\end{figure}

We consider four networks for each of the 100 Facebook data sets: the largest connected component of the full network (which we hereafter identify as ``Full"), the largest connected component of the student-only network (``Student"), the largest connected component of the female-only network (``Female"), and the largest connected component of the male-only network (``Male").  The Male and Female networks are each subsets of the Full network rather than the Student network.  Each network has a single type of unweighted, undirected connection between nodes and can thus be represented as an adjacency matrix $\mathbf{A}$ with elements $A_{ij}=A_{ji}$ indicating the presence ($A_{ij} = 1$) or absence ($A_{ij} = 0$) of a tie between nodes $i$ and $j$. The resulting tangle of nodes and links, which we illustrate for the Reed College student Facebook network in Figure \ref{viz}, can obfuscate any organizational structure that might be present. 

The data also includes limited demographic (categorical) information that is volunteered by users on their individual pages: gender, class year, and (using anonymous numerical identifiers) high school, major, and residence.  We use a ``Missing'' label for situations in which individuals did not volunteer a particular characteristic. The different characteristics allow us to make comparisons between institutions, under the assumption (see the discussion by \cite{boydell}) that the communities and other elements of structural organization in Facebook networks reflect (even if imperfectly) the social communities and organization of the offline networks on which they're based.  It is an important research issue to determine just how imperfect this might be \citep{hogan09}, but this is far beyond the scope of the present paper (though we hope that others will take on this particular challenge).  The conclusions that we draw in this paper apply directly to the university Facebook networks from September 2005, and we expect that they can provide insight about the real-world social networks at the institutions as well.

%%%%%%%%%%

\section{Methods}

We study each network at both the dyad level and the community level.  We first consider homophily \citep{faust,miller01,newman2010}) quantified by assortativity coefficients using the available categorical data.  For some of the smaller networks, we additionally perform independent logistic regression on node pairs to obtain the log odds contributions to edge presence between two nodes that have the same categorical-data value.  We similarly fit exponential random graph models (ERGMs) \citep{ergm, RobinsERGM,fs86,wp96,lubbers07} with triangle terms to these smaller networks.  Finally, we partition the networks by algorithmically detecting communities \citep{ourreview,santobig}, which we compare to the given categorical data using the technique in this paper's prequel \citep{siamface}.  Calculating assortativity values and log odds contributions allows us to examine ``microscopic'' features of the networks, while comparing algorithmic partitions of the networks to the categorical data allows us to examine their ``macroscopic'' features.  As we illustrate below, both perspectives are important because they provide complementary insights.

%%%%%%%%%%%%%%
   
\subsection{Assortativity}

A general measure of scalar assortativity $r$ relative to a categorical variable is given by \cite{newman_2003,newman2010}:
\begin{equation}
	r = \frac{\mbox{tr} ({\bf e}) - \| {\bf e}^2 \|}{1 - \| {\bf e} ^2 \|} \in [-1,1] \,,
\end{equation}
where ${\bf e} = {\bf E }/\| {\bf E} \|$ is the normalized mixing matrix, the elements $E_{ij}$ indicate the number of edges in the network that connect a node of type $i$ (e.g., a person with a given major) to a node of type $j$, and the entry-wise matrix $1$-norm $\| {\bf E } \|$ is equal to the sum of all entries of ${\bf E}$.  By construction, this formula yields $r=0$ when the amount of assortative mixing is the same as that expected independently at random (i.e., $e_{ij}$ is simply the product of the fraction of nodes of type $i$ and the fraction of nodes of type $j$), and it yields $r=1$ when the mixing is perfectly assortative.

%%%%%%%%%

\subsection{Logistic Regression and Exponential Random Graphs}
\label{sec:regress}

We further measure the influence of the available user characteristics on the likelihood of a ``friendship'' tie via a fit by logistic regression (under an assumption of independent dyads) and by an ERGM specification that includes triangle terms.  Our focus is on trying to calculate the propensity for two nodes with the same categorical value to form a tie.  We consider each of the four categorical variables (major, residence, year, and high school) and use the {\tt ERGM} package in R \citep{ergm} for both models (treating each network as undirected).  We used R 2.11.1 and the {\tt statnet} package version 2.1-1, and we note that different versions of R and {\tt statnet} caused different degrees of convergence with the structural elements in the model.  We obtained results for the 16 smallest institutions.  (We did these calculations on a 32-bit operating system, which restricts the network sizes that can be processed.) Both models that we consider are based on a standard ERGM parametrization
$P_\mathbf{\theta}\{\mathbf{Y}=\mathbf{A}\} = \exp\{\mathbf{\theta}\cdot\mathbf{g}(\mathbf{A})\} / \kappa(\mathbf{\theta})$ describing the distribution of graphs with model coefficients $\mathbf{\theta}$ corresponding to statistics calculated from the adjacency matrix $\mathbf{A}$ (with a normalizing factor $\kappa$ to ensure that the formula yields a probability distribution) \citep{ergm, RobinsERGM,fs86,wp96,lubbers07}. 

In the first model (logistic regression), we include five statistics (with five corresponding $\mathbf{\theta}$ coefficients): the total density of ties ({\tt edges}) and the common classifications ({\tt nodematch}) from each of four node/user characteristics: residence, class year, major, and high school.  For example, the $\theta_\mathrm{high school}$ contribution describes the additional log-odds predisposition for a ``friendship" tie when two users are from the same high school. In all cases, we ignore possible contributions from missing characteristic data: two nodes with the same missing data field are not treated as having the same value for the characteristic.  Rather than include gender explicitly in the model, we instead additionally fit the model to the single-gender subnetworks in order to be consistent with the treatment of gender in the community-level comparisons below. In the second model (an ERGM), we add a {\tt triangle} statistic to account for the observed amount of transitivity in the network data.  This gives a total of six $\mathbf{\theta}$ coefficients: edges, common residence, common class year, common major, common high school, and the triangle coefficient.

%%%%%%%%

\subsection{Community Detection}

The global organization of social networks often includes coexisting
modular (horizontal) and hierarchical (vertical) organizational
structures, and myriad papers have attempted to interpret such
organization through the computational identification of ``community
structure." Communities are defined in terms
of cohesive groups of nodes with more internal connections (between
nodes in the same group) than external connections (between nodes in the
group and nodes in other groups).  As discussed at length in two
recent review articles \citep{ourreview,santobig} and in references
therein, the ensemble of techniques available to detect communities is
both numerous and diverse.  Existing techniques include hierarchical clustering
methods such as single linkage clustering, centrality-based methods,
local methods, optimization of quality functions such as modularity
and similar quantities, spectral partitioning, likelihood-based
methods, and more.  Communities
%, which are not typically identified in advance, 
are considered to not be merely
structural modules but are also expected to have functional importance
because of the large number of common ties among nodes in a community.
For example, communities in social networks might correspond to
circles of friends or business associates and communities in the World
Wide Web might encompass pages on closely-related topics.
In addition to remarkable successes on benchmark
problems, investigations of community structure have
observed correspondence between communities and ``ground truth''
groups in diverse application areas---including the reconstruction
of college football conferences \citep{structpnas} and the
investigation of such structures in algorithmic rankings \citep{bcs};
the investigation of committee assignments \citep{congshort}, legislation
cosponsorship \citep{yan}, and voting blocs \citep{waugh,multislice} in the
United States Congress; the examination of functional groups in metabolic networks \citep{amaral}; the study of ethnic preferences in school friendship networks \citep{marta}; and the study of social structures in mobile-phone conversation networks \citep{jp}

\begin{figure}[htbp]
\centerline{
\includegraphics[width=.6\textwidth]{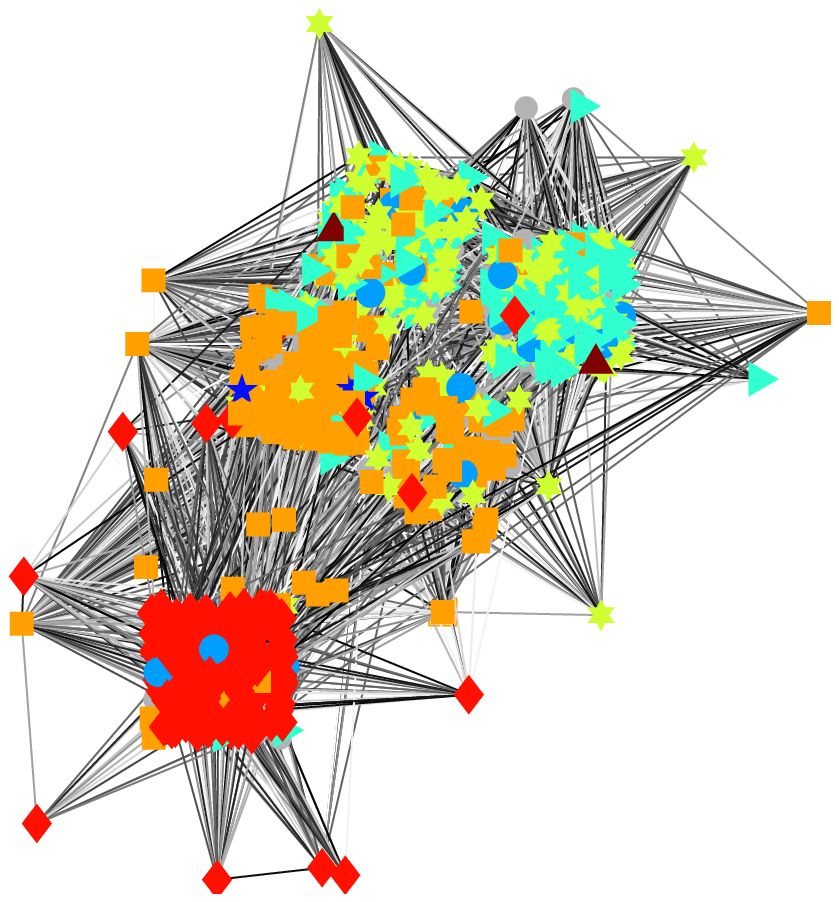}
\includegraphics[width=.4\textwidth]{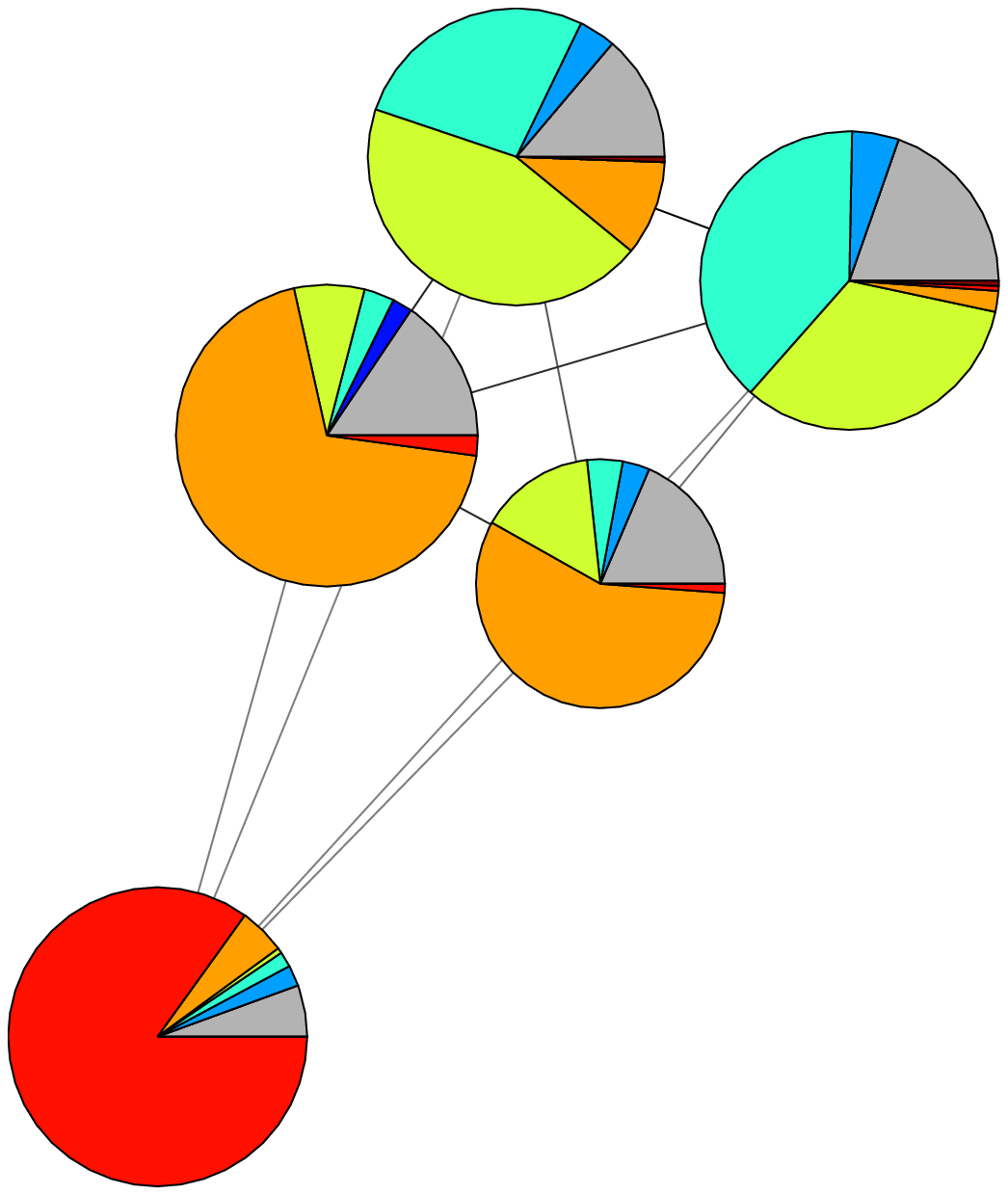}
}
\caption{[Color] (Left) Vizualization of community structure of the Reed College Student Facebook network shown in Figure \ref{viz}.  Node shapes and colors indicate class year (gray dots denote users who did not identify an affiliation), and the edges are randomly shaded for easy viewing.  We place the communities using a \cite{FruchtermanReingold91} layout and use a \cite{kk} layout to position the nodes within communities \citep{viz09}. (Right) The same network layout but with each community depicted as a pie.  Larger pies represent communities with larger numbers of nodes.  Darker edges indicate the presence of more connections between the corresponding communities.} 
\label{vizcommdetect}
\end{figure}

In the present paper, we investigate the community structures of the Facebook networks
from each of the 100 colleges and universities. (See the visualization of the
community structure for Reed College in Figure \ref{vizcommdetect}.)
For each institution, we consider the Full, Student, Female, and Male
networks.  We seek to determine how well the demographic labels
included in the data correspond to algorithmically computed communities.
Assortativity provides a local measure of homophily, but that does not
provide sufficient information to draw conclusions about the global
organization of the Facebook networks.  For example, two students who
attended the same high school are typically more likely to be friends
with each other than are two students who attended different high
schools, but this will not necessarily have a meaningful
community-level effect unless enough of the
students went to common high schools.  As we we will see
below, high school tends to be a much more dominant organizing
characteristic of the social structure at the large institutions
than at small institutions, presumably because of a significant frequency of common high school pairs at the large institutions.

We identify communities by optimizing the ``modularity'' quality
function $Q = \sum_i (e_{ii} - b_i^2)$,
where $e_{ij}$ denotes the fraction of ends of edges in group~$i$ for
which the other end of the edge lies in group~$j$ and $b_i = \sum_j
e_{ij}$ is the fraction of all ends of edges that lie in
group~$i$. High values of modularity correspond to community
assignments with greater numbers of intra-community links than
expected at random (with respect to a particular null model
\citep{newman2006pre,ourreview,santobig}).  Although numerous other community
detection methods are also available, modularity optimization
is perhaps the most popular way to detect communities and it has been
successfully applied to many applications \citep{ourreview,santobig}.
One might also consider using a method that includes a resolution
parameter \citep{spinglass} to avoid issues with resolution limits
\citep{resolution}.  However, our primary focus is on global organization of the networks, so we limit our attention to the default resolution of modularity. This focus arguably biases our study of communities to the largest structures, such as those influenced by common class year, making the observed correlations with other demographic characteristics even more striking.

To try to ensure that the communities we detect are properties of the
data rather than of the algorithms that we used, we optimize modularity (with default resolution) using 6 different combinations
%\footnote{\bf I deliberately changed this from
%8 to 6 because I cannot see from the description how the ``double''
%and ``both'' should be different from one another without different information in their descriptions, we talked about this.} 
of spectral optimization, greedy optimization,
and \cite{kl} (KL) node-swapping steps (in the manner discussed 
by \cite{newman2006pnas}).  
Specifically, we use (1) recursive
partitioning by the leading eigenvector of a modularity matrix
\citep{newman2006pre}, (2) recursive partitioning by the leading pair of
eigenvectors (including the \cite{richardson} extension of the method in \cite{newman2006pre}), (3) the Louvain greedy method \citep{greedy}, and each
of these three supplemented with small increases in the quality $Q$ that can be obtained using KL node swaps.  Each of these 6 methods yields a community partition, and we obtain our comparisons (described in Section \ref{sec:CompareComm}) by considering each of these 6 partitions.

Modularity optimization is NP-hard \citep{np}, so one must be cautious
about the large number of degenerate partitions in the modularity
landscape \citep{good2010}.  However, by detecting coarse
observables---in particular, the global organization of a Facebook
network based on the given categorical data---and considering results
that are averaged over multiple optimization methods, one can obtain
interesting insights.  The specific ``best" partition will vary from
one method to another, but some of the predicted coarse
organizational structure of the networks (see below) is robust to the choice of community detection
algorithm.

%%%%%%%%%

\subsection{Comparing Communities to Node Data} \label{sec:CompareComm}

Once we have detected communities for each institution, we will
compare the algorithmically-obtained community structure to the
available categorical data for the nodes.  We recently developed a
methodology to accomplish this goal in \cite{siamface} (where we
considered only 5 institutions among the 100 in order to illustrate
the techniques). This method of comparison can be applied to the output of any ``hard partitioning'' algorithm in which each node is assigned to precisely one community (cf. ``soft partitioning'' methods, in which communities can overlap).
We briefly review that methodology here. 

To compare a network partition to the categorical demographic data, we standardize (using a $z$-score) the Rand coefficient of the communities in that partition compared to partitioning based purely on each of the
four categorical variables (one at a time).  For each
comparisons, we calculate the Rand $z$-score $z$ in terms of the total number
of pairs of nodes in the network $M$, the number of pairs that are in
the same community $M_1$, the number of pairs that have the same
categorical value $M_2$, and the number of pairs of nodes that are
both in the same community and have the same categorical value $w$
\citep{siamface}. The Rand coefficient is given in term of these
quantities by $S = [w+(M-M_1-M_2+w)]/M$ \citep{Rand71}.  We then calculate
the $z$-score for the Rand coefficient as \citep{Hubert77,siamface}
\begin{equation}\label{eq:zrand}
	z = \frac{1}{\sigma_w}\left(w-\frac{M_1M_2}{M}\right)\,,
\end{equation}
where
\begin{align}
	\label{eq:varw}
	\sigma_w^2 &= \frac{M}{16} - \frac{(4M_1-2M)^2(4M_2-2M)^2}{256M^2} + \frac{C_1C_2}{16n(n-1)(n-2)}\cr &+ \frac{[(4M_1-2M)^2-4C_1-4M][(4M_2-2M)^2-4C_2-4M]}{64n(n-1)(n-2)(n-3)}\,,
\end{align}
$n$ is the number of nodes in the network, the coefficients $C_1$ and $C_2$ are given by
\begin{align} \label{eq:C12}
	C_1 &= n(n^2-3n-2) - 8(n+1)M_1 + 4\sum_i n_{i\cdot}^3\,,\cr
	C_2 &= n(n^2-3n-2) - 8(n+1)M_2 + 4\sum_j n_{\cdot j}^3\,,
\end{align}
$n_{ij}$ denotes an element of a contingency table and indicates the number of nodes that are classified into the $i$th group of the first partition and the $j$th group of the second partition, $n_{i\cdot}=\sum_j n_{ij}$ is a row sum, and $n_{\cdot j}=\sum_i n_{ij}$ is a column sum.  Each $z$-score indicates the deviation from randomness in comparing the community structure with the partitioning based purely on that single demographic characteristic. One needs to be cautious when interpreting such deviations from randomness as a strength of correlation.  In particular, given the dependence on system size inherent in this measure, one should not overinterpret the relative values of $z$-scores from different institutions. Nevertheless, the $z$-scores provide a reasonable proxy quantity both for the statistical significance of correlation and for the relative strength of correlation in a specified network.

%%%%%%%%

\section{Results}

We now use the methods outlined in the previous section to study the Facebook networks.  We first follow the order of presentation above and then make some observations in combinations. Complete results are available in the tables in the appendix.

%%%%%

\subsection{Assortativity}

We tabulate the assortativities based on gender, major, residence, class
year, and high school for all networks (and subsets thereof) in Table
\ref{tab:AssortTable}.  

For almost all of the institutions and each of the 4 network subsets, the class
year attribute produces higher assortativity values than the other
available demographic characteristics.  However, Rice University (31), California Institute of Technology (36), University of Georgia (50), University of Michigan (67), Auburn University (71), and University of Oklahoma (97) are each examples in which residence provides the highest
assortativity values (again, for each of the 4 network subsets).
We discussed Caltech as a focal example in
\cite{siamface}, in which we introduced the community comparison
methods that we employ below.  

Other institutions have varying orderings of
class year and residence assortativity among the 4 network
subsets.  At MIT (8), USF (51), Notre Dame (57), University of
Maine (59), UC (61), UC (64), and MU (78), residence gives the highest
assortativity in the Male networks.  The UCF (52) Female network
has its highest assortativity with residence.  Both the Full network
and the Male network for University of California at Santa Cruz
(68) have their highest assortativity values with residence.  Both
the Male and Female networks at University of Illinois at
Urbana-Champaign (20), Tulane (29), UC (33), Florida State University
(53), Cal (65), University of Mississippi (66), University of Indiana
(69), Texas (80), Texas (84), University of Wisconsin (87), Baylor
(93), University of Pennsylvania (94), and University of Tennessee
(95) have their highest assortativity values with residence; all
other networks from these institutions have their highest
assortativity with class year.  

Some outlying observations can be tied directly to small samples. For example, Simmons (81) is a female-only college.  It has only four males in the Full network; none of the males had any connections with another male, so the gender assortativity values for both the Full and Student components are very close to $0$. Similar gender numbers are also present in the data from Wellesley (22) and Smith (60).

%%%%%%%

\subsection{Dyad-Level Regression and Exponential Random Graphs}

We use the two statistical models described in Section \ref{sec:regress} to study the 16 smallest institutions.  The (dyad-independent) logistic regression model includes contributions from edges (network density) and matched user (node) characteristics for each of four demographic variables.  We present the results for this model in Table \ref{tab:LogOdds2}. The second model that we consider
is an ERGM, which supplements the first model with a structural {\tt
triangle} contribution.  We present the results for this model in Table
\ref{tab:LogOdds1}. These calculations give views of the networks at the microscropic (dyad-level) scale that supplement the results that we obtained using the assortativity statistics.

\begin{figure}[htbp]
\centerline{
\includegraphics[width=.5\textwidth]{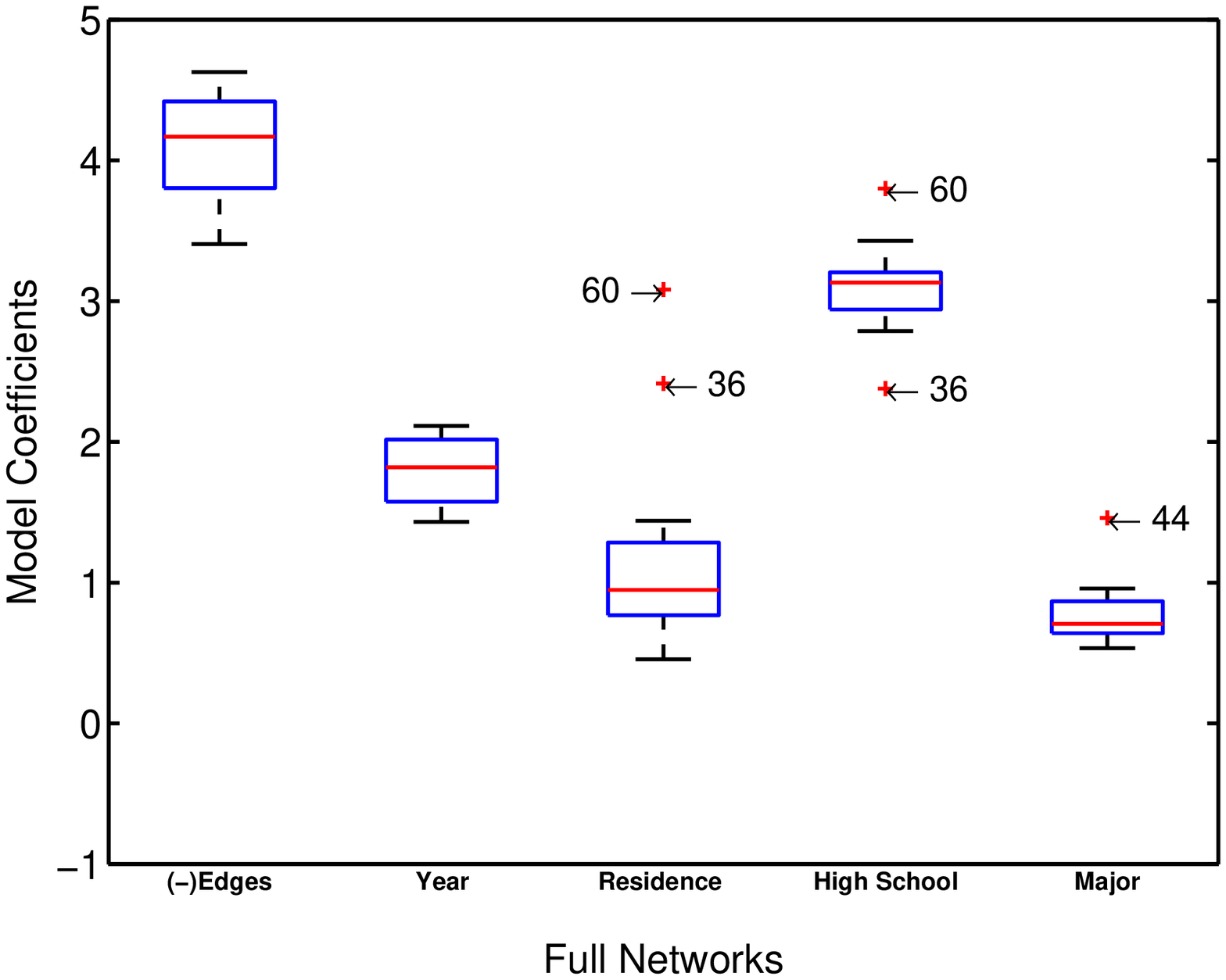}
\includegraphics[width=.5\textwidth]{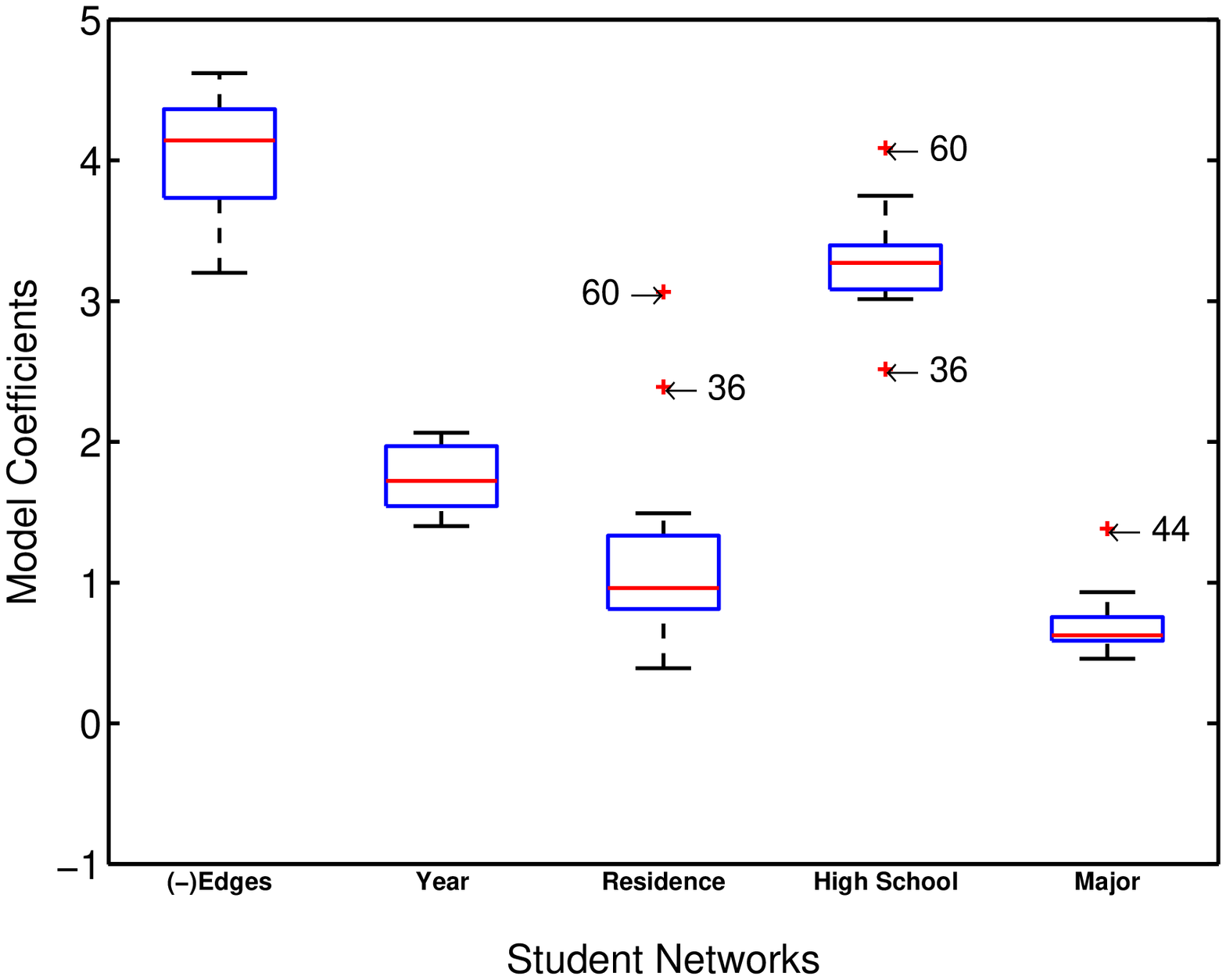}}
\centerline{\includegraphics[width=.5\textwidth]{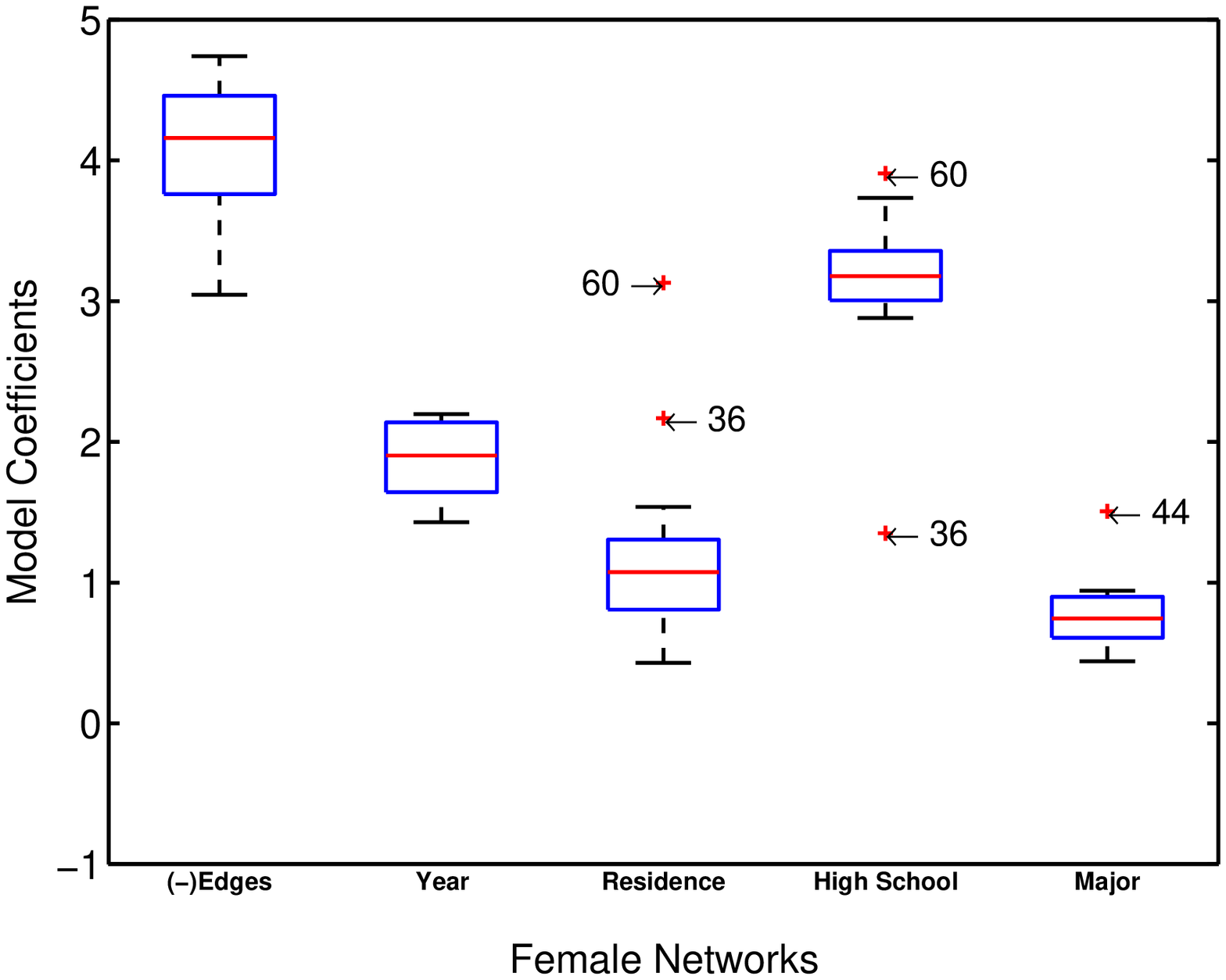}
\includegraphics[width=.5\textwidth]{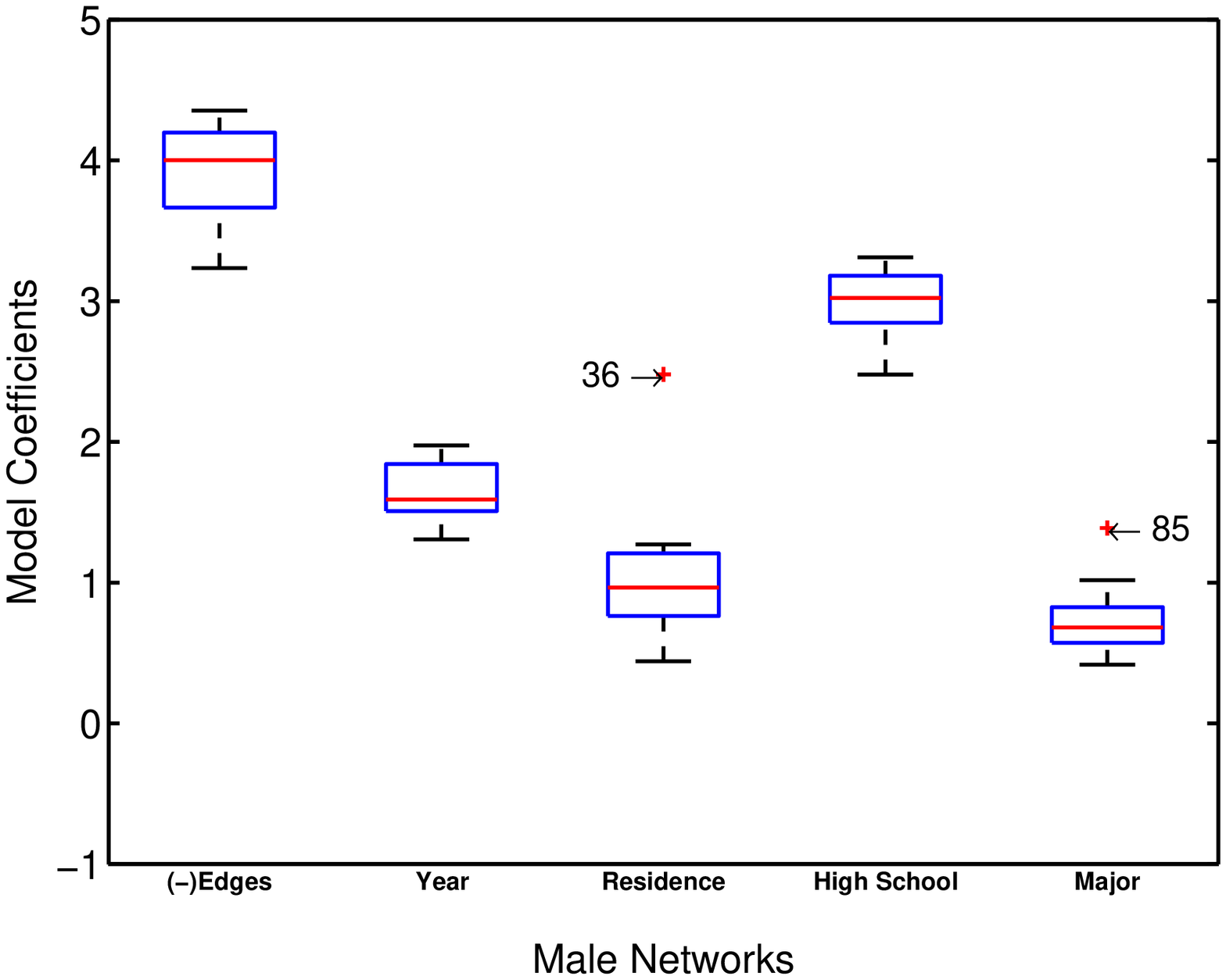}}
\caption{Box plots (indicating median, quartiles, extent, and outliers of the distribution) of the logistic regression {\tt nodematch} coefficients for the 16 smallest institutions in the data for the model described in the main text. We plot the $-\theta_\mathrm{edges}$ values to present results with greater resolution.  We separately present our results for the Full, Student, Female, and Male networks.}
\label{fig:BoxPlots}
\end{figure}

\begin{figure}[htbp]
\centerline{
\includegraphics[width=.5\textwidth]{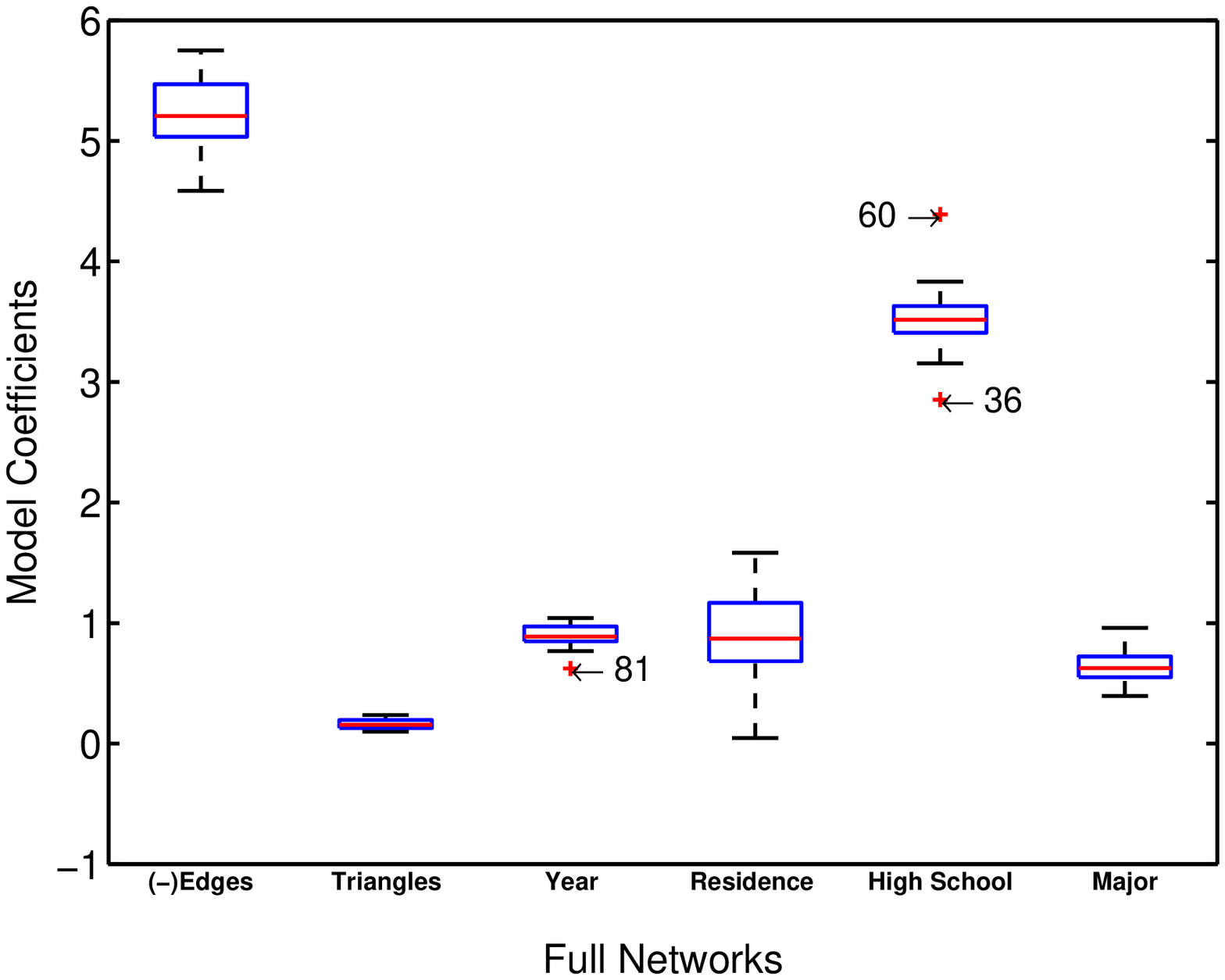}
\includegraphics[width=.5\textwidth]{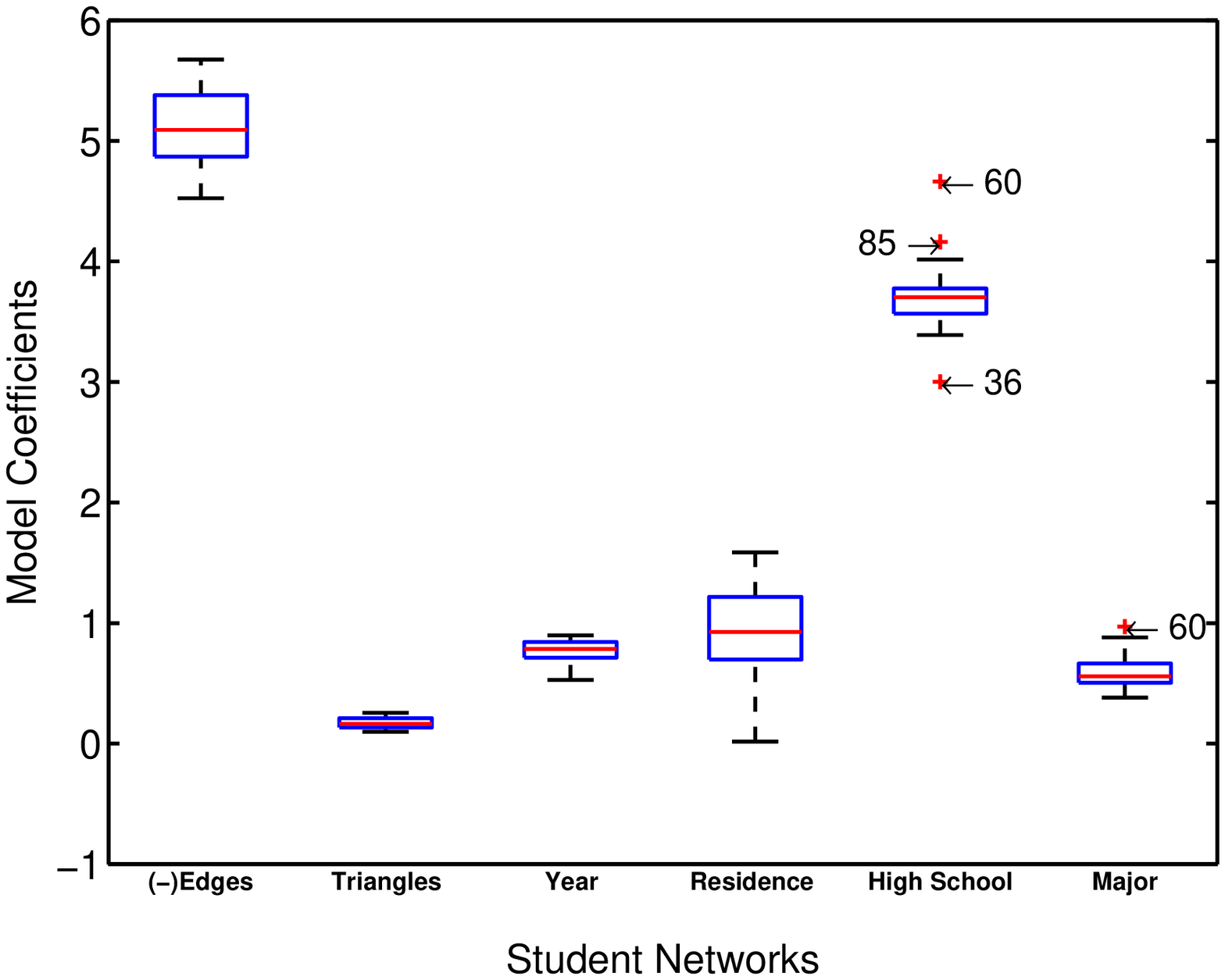}}
\centerline{\includegraphics[width=.5\textwidth]{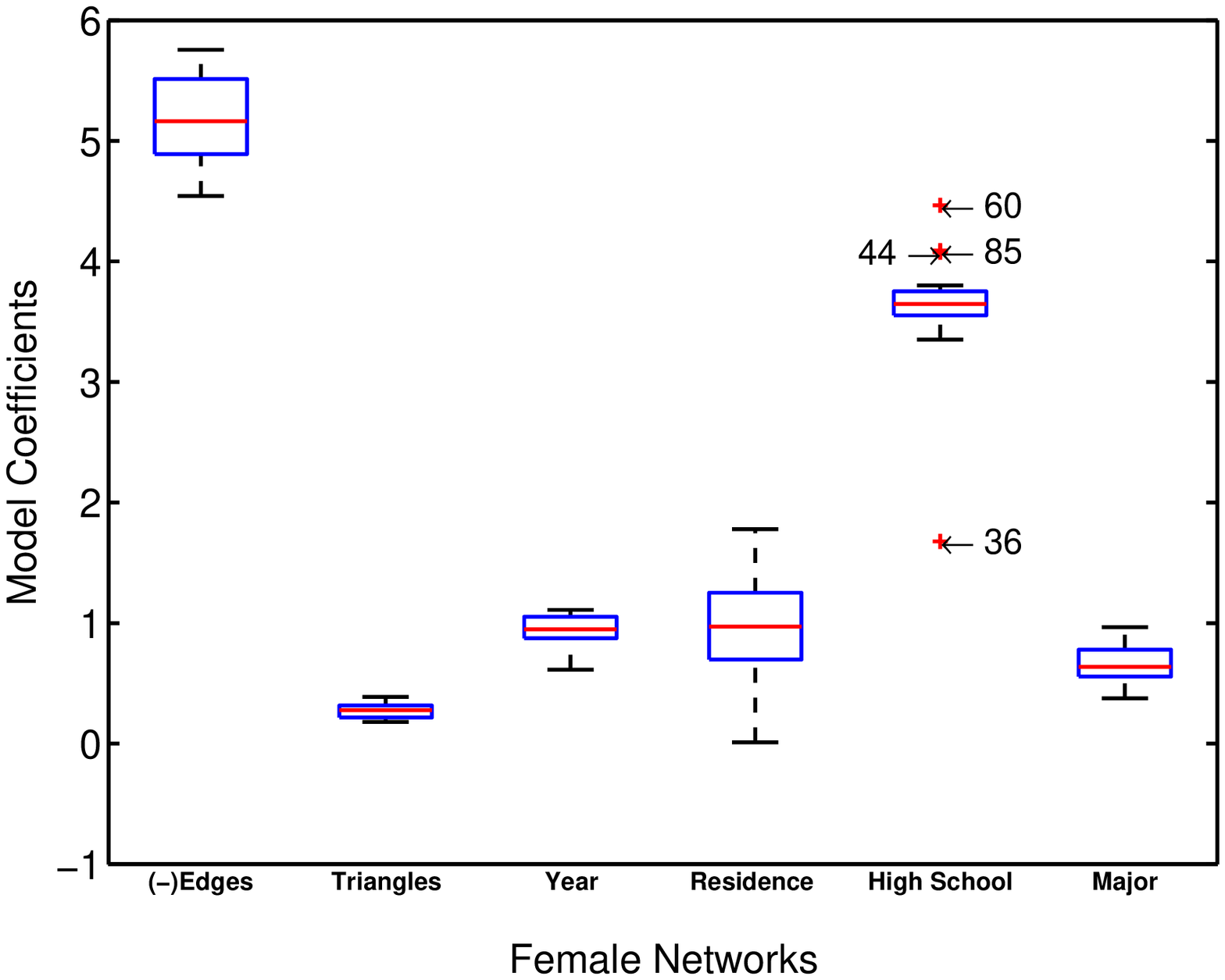}
\includegraphics[width=.5\textwidth]{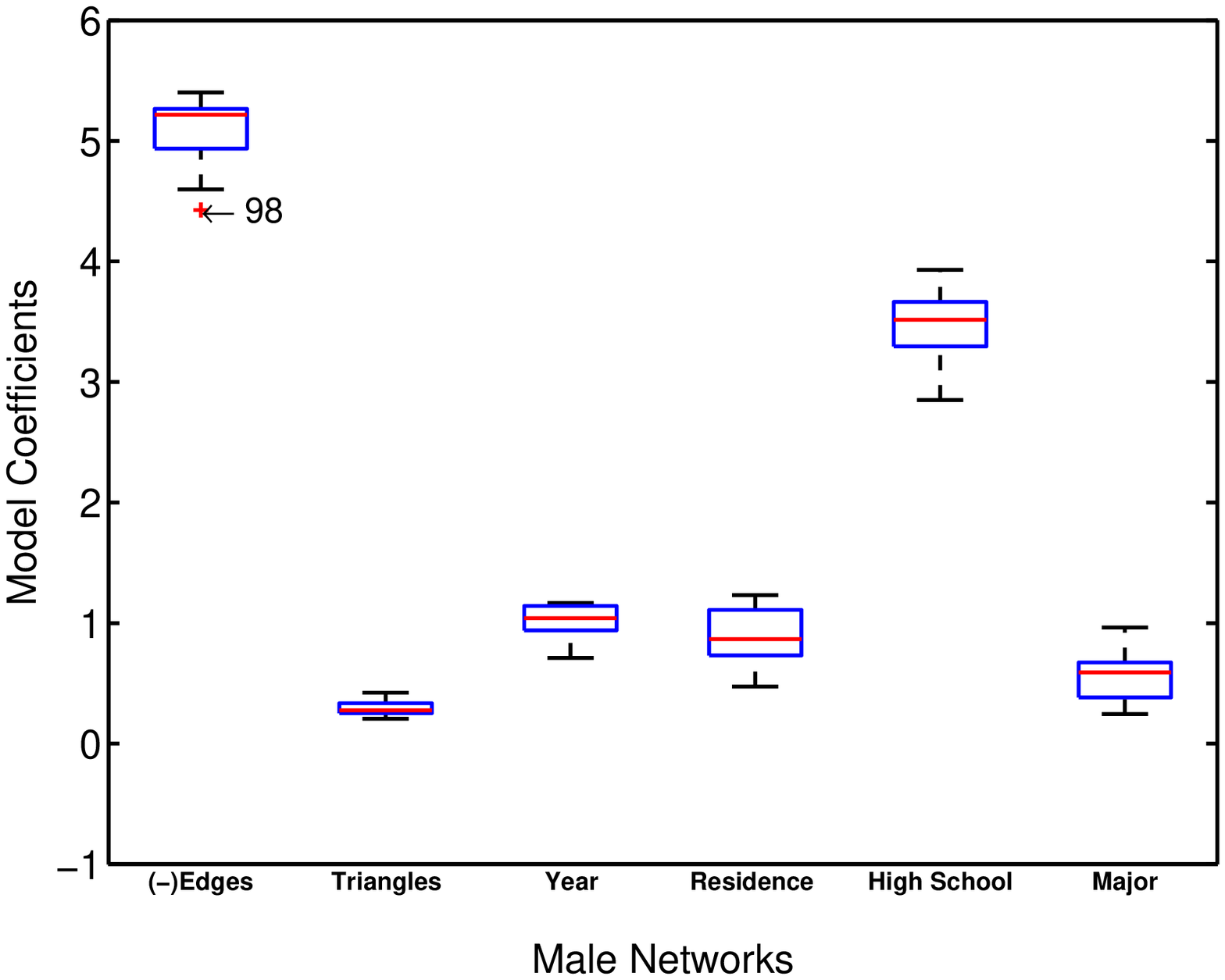}}
\caption{Box plots (indicating median, quartiles, extent, and outliers of the distribution) of the exponential random graph model coefficients described in the main text for the 16 smallest institutions in the data. We plot the $-\theta_\mathrm{edges}$ values to present results with greater resolution. We separately present our results for the Full, Student, Female, and Male networks.}
\label{fig:TriBoxPlots}
\end{figure}

We consider the results from the 16 smallest institutions by fitting the models to each of their Full, Student, Female, and Male networks. Because all of the resulting model coefficients appear to be statistically significant at a $p$-value of less than $10^{-4}$, we interpret the importance of node matching on the different demographic characteristics directly from the magnitude of the corresponding model coefficients. We summarize the results for these 16 institutions using the box plots in Figures \ref{fig:BoxPlots} and \ref{fig:TriBoxPlots}. The box plots identify the outliers by institution number: Caltech (36), Oberlin (44), Smith (60), Simmons (81), Vassar (85), and Reed (98). (As we have only performed this regression analysis for the 16 smallest institutions in the data, one should not jump to conclusions from this list of outliers.)
For all institutions and all four types of networks for each institution, the highest coefficient in the employed ERGM model (with {\tt triangle} terms) is given for matching the High School category, and the value of this coefficient is significantly higher than those for the other node-matching coefficients.  Only the Caltech (36) Female network has ERGM  coefficients for Year,
Residence, and High School that are very close to each other.  

%%%%%%%%%

\subsection{Comparison of Communities}

We now discuss community-level results for each network using $z$-scores of the Rand coefficient to compare partitions obtained via algorithmic community detection to partitions based on each characteristic. That is, each community-detection result identifies a group assignment for each node, thereby producing a partition (called a ``hard" partition) in which each node is assigned to exactly one community.  One can also obtain a hard partition for each network by selecting a single characteristic and grouping nodes according to that characteristic. Every network that we study (including the subnetworks) has at least one $z$-score in the set $\{z_{\mbox{Major}},z_{\mbox{Year}},z_{\mbox{HS}},z_{\mbox{Residence}}\}$ with a value greater than $5$.  Although the distribution of Rand coefficients is decidedly not Gaussian, particularly in the tails of the distributions \citep{siamface,BrookStirling84,Kulisnkaya}, this 
%five standard-deviation 
$z = 5$
threshold indicates that at least one characteristic in each network exhibits strong statistical significance. Moreover, we will see that the vast majority of our comparisons below exceed the 
$z = 2$
%two standard-deviation 
threshold. (That is, they essentially lie outside 95\% confidence intervals.)

To visualize and compare the varied strengths of organization according to the different demographic characteristics, we represent the four $z$-scores obtained for each network (Full, Student, Female, and Male) of an institution using 3-dimensional barycentric (tetrahedral) coordinates \citep{mathworld,jnf}.  We start by setting all negative $z$-scores to $0$, as all observed negative $z$-score values are small enough to be statistically insignificant.  We then normalize by the sum of the $z$-scores to obtain
\begin{align}
	z_{1}&=\frac{z_{\mbox{Major}}}{z_{\mbox{Major}}+z_{\mbox{Year}}+z_{\mbox{HS}}+z_{\mbox{Residence}}} \,,\cr
	z_{2}&=\frac{z_{\mbox{Residence}}}{z_{\mbox{Major}}+z_{\mbox{Year}}+z_{\mbox{HS}}+z_{\mbox{Residence}}}\,,\cr
	z_{3}&=\frac{z_{\mbox{Year}}}{z_{\mbox{Major}}+z_{\mbox{Year}}+z_{\mbox{HS}}+z_{\mbox{Residence}}}\,,\cr
	z_{4}&=\frac{z_{\mbox{HS}}}{z_{\mbox{Major}}+z_{\mbox{Year}}+z_{\mbox{HS}}+z_{\mbox{Residence}}}\,.
\label{eq:NormZ}
\end{align}
From these 4 $z$-score values, we calculate coordinates $X = (x_1,x_2,x_3)$ located inside a tetrahedron.  For example, one can obtain a tetrahedron whose vertices are $p_{1}=(1,0,0)$, $p_{2}=(\cos(2\pi/3),\sin(2\pi/3),0)$, $p_{3}=(\cos(4\pi/3), \sin(4\pi/3), 0)$, and $p_{4}=(0,0,\sqrt{2})$) with the transformation
\begin{align}
	X &= (T \times Z)+p_{4} \,, \notag \\
	T &= \left[\begin{array}{ccc} p_{1}-p_{4} & p_{2}-p_{4} & p_{3}-p_{4} \end{array} \right] \,, \notag \\
	Z &= \left[\begin{array}{c} z_1 \\ z_2 \\ z_3 \end{array}\right]\,.\label{grrrrr}
\end{align}
The information from $z_4 = 1 - (z_1 + z_2 + z_3)$ is implicitly included in (\ref{grrrrr})
because of the normalization. Each of the 4 vertices of the tetrahedron corresponds to a limit in which the corresponding $z$-score completely dominates the other three $z$-scores.  That is, at a vertex, the entire $z$-score sum arises from the corresponding component.

\begin{figure}[htbp]
\vspace*{-0.25in}\hspace*{-0.75in}{\includegraphics[width=.85\textwidth]{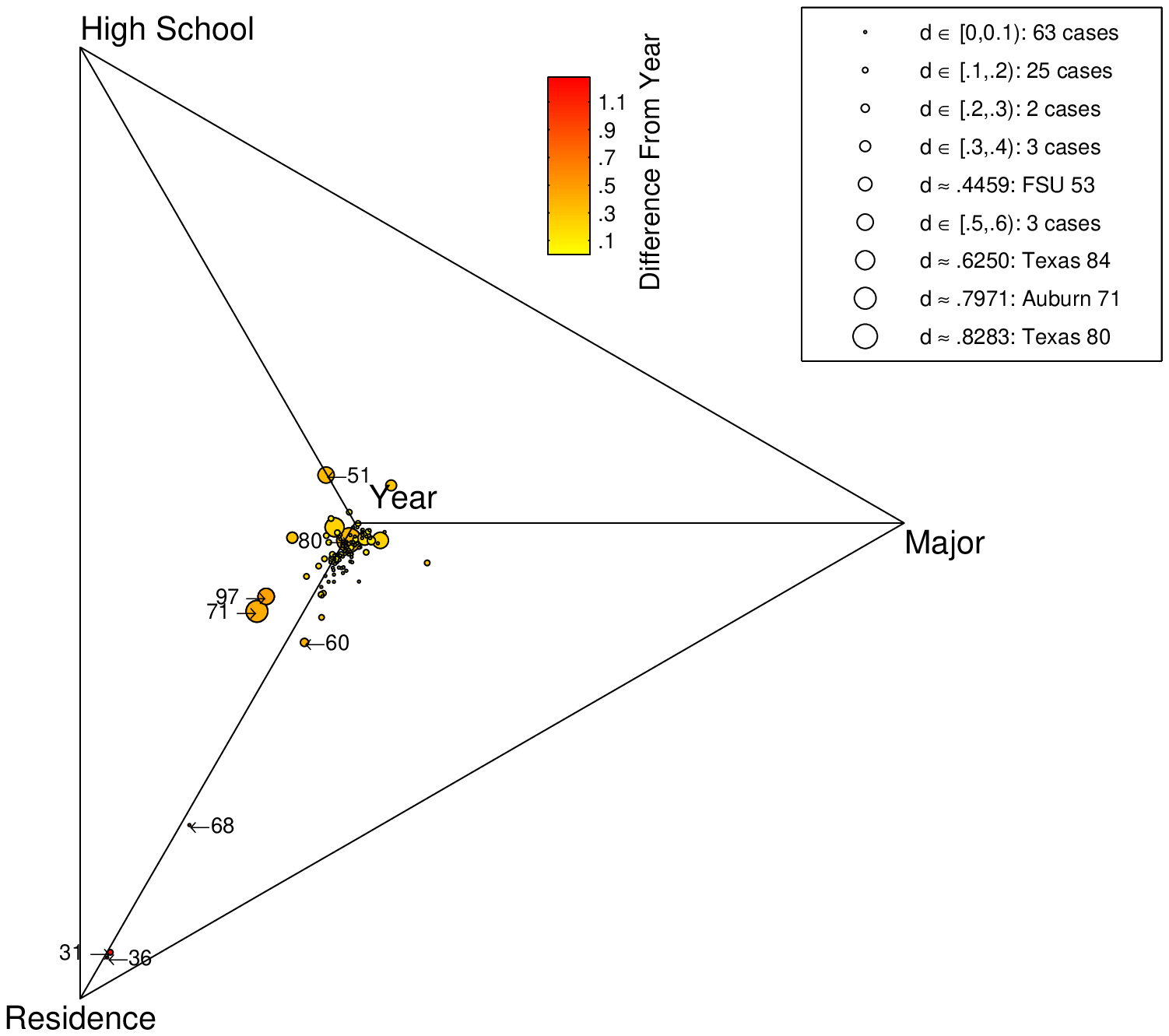}}
\vspace*{-1.05in}\\
\hspace*{1.5in}{\includegraphics[width=.85\textwidth]{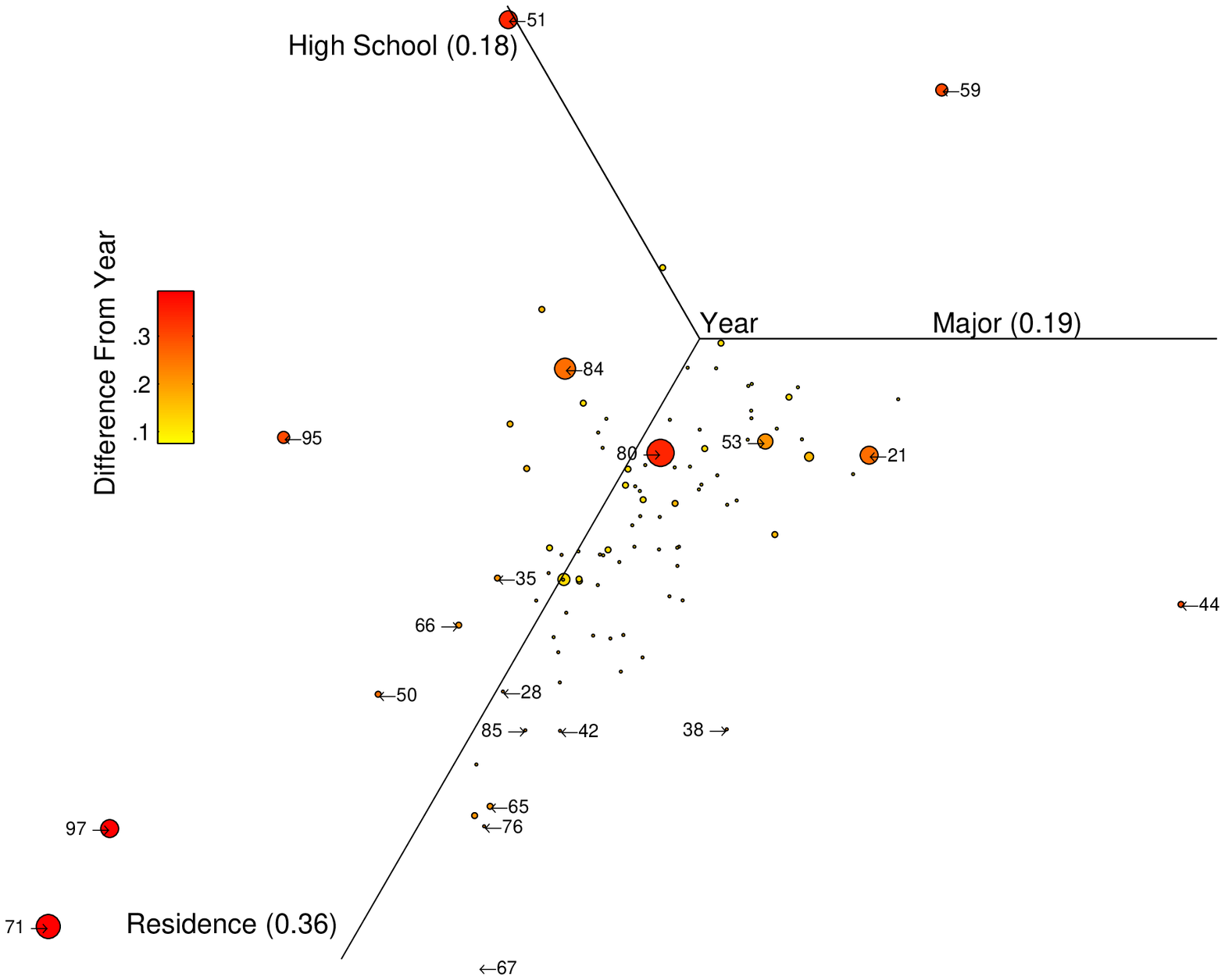}}
\caption{[Color online] (Upper Left) Social organization tetrahedron for the community structures of the Full component (largest connected component) of the networks for each of the 100 institutions.  Lighter disks indicate an organization that is based more predominantly on class year.  See the main text for a description of this figure.  (Lower Right) Magnification near the Year vertex. The legend illustrates the disk size as a function of the maximum distance $d$ between the 6 different partitions of the network.  Most cases (88 out of 100 institutions) have $d<.2$.
%The legend illustrates the disk size as a function of the maximum distance, $d$, between the six different partitions of the network.  FSU 53 has a maximum distance of $.4459$ between the first partition, the recursive Leading Eigenvector Method, and the third partition, greedy optimization.  Most distances fall within $[0,.1)$, 72 cases out of the total 100 institutions.  For those cases where $d>.4$, the maximum distance for three out of the seven cases is between partitions 1 and 3, UF 21, FSU 53, and Texas 84, for three more cases the maximum distance is between partition 1 and partition 6 (greedy optimization with KL node swaps), Auburn 71, Texas 80, and Oklahoma 97, and the last case, USF 51, has the largest distance between partition 2 and partition 6.
}
\label{fig:All}
\end{figure}

\begin{figure}[htbp]
\vspace*{-0.25in}\hspace*{-0.75in}{\includegraphics[width=.85\textwidth]{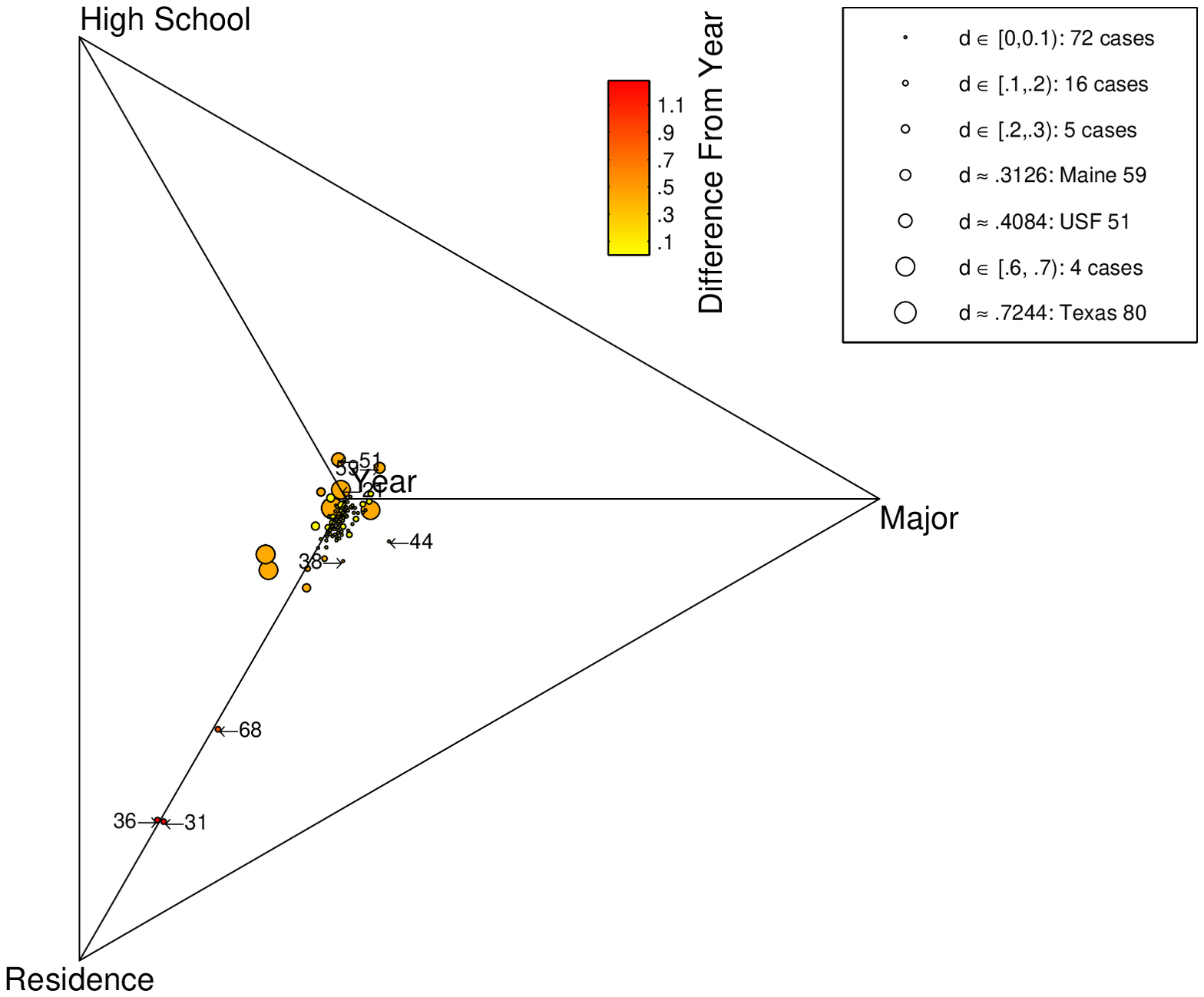}}
\vspace*{-1.05in}\\
\hspace*{1.5in}{\includegraphics[width=.85\textwidth]{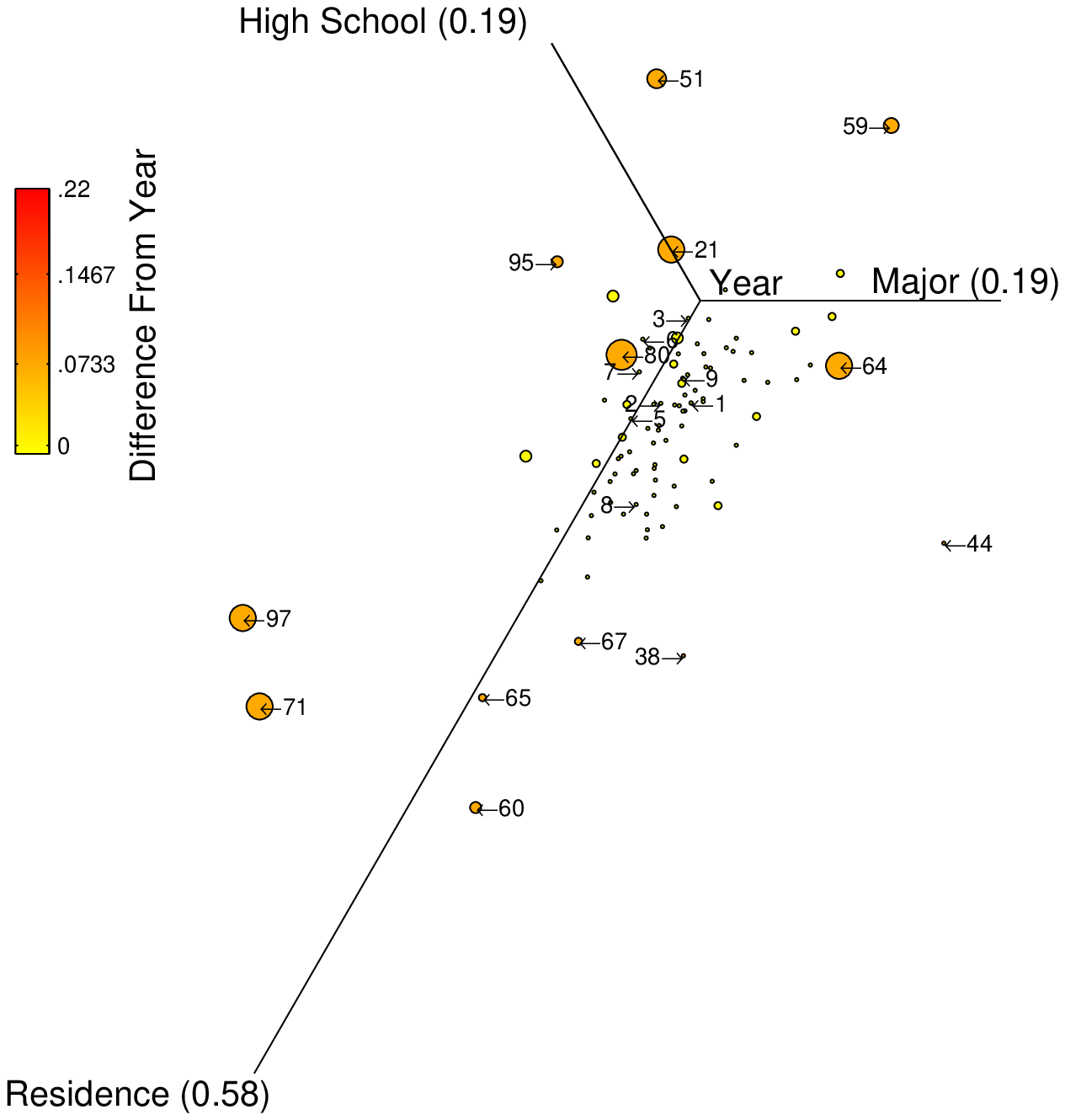}}
\caption{[Color online] (Upper Left) Social organization tetrahedron for the community structures of the Student component of the networks for each of the 100 institutions.  Lighter disks indicate an organization that is based more predominantly on class year.  See the main text for a description of this figure.  (Lower Right) Magnification near the Year vertex.  As in Figure \ref{fig:All}, the disk sizes correspond to the maximum distances between partitions.
%For this diagram, there are six cases in which the maximum distance between partitions exceeds $.4$.  Two of these cases have the maximum distance between partitions 1 and 6,USF 51 and Texas 80, two of these cases have the maximum distance between partitions 2 and 6, UF 21 and Oklahoma 97, one case has its maximum distance between partitions 1 and 3, UC 64, and the last case has its maximum distance between partitions 2 and 4, Auburn 71.
}
\label{fig:Student}
\end{figure}

\begin{figure}[htbp]
\vspace*{-0.25in}\hspace*{-0.75in}{\includegraphics[width=.85\textwidth]{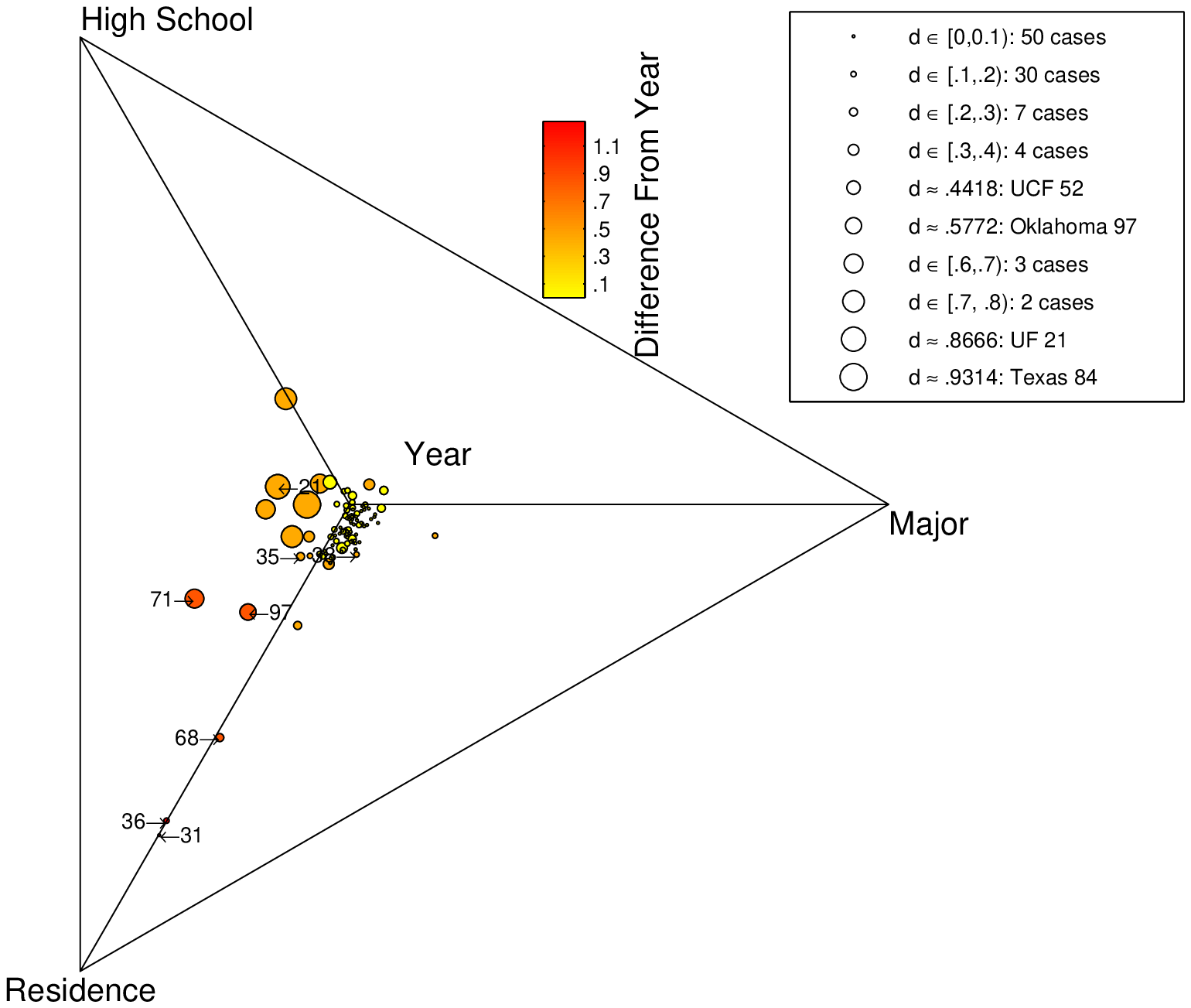}}
\vspace*{-1.05in}\\
\hspace*{1.5in}{\includegraphics[width=.85\textwidth]{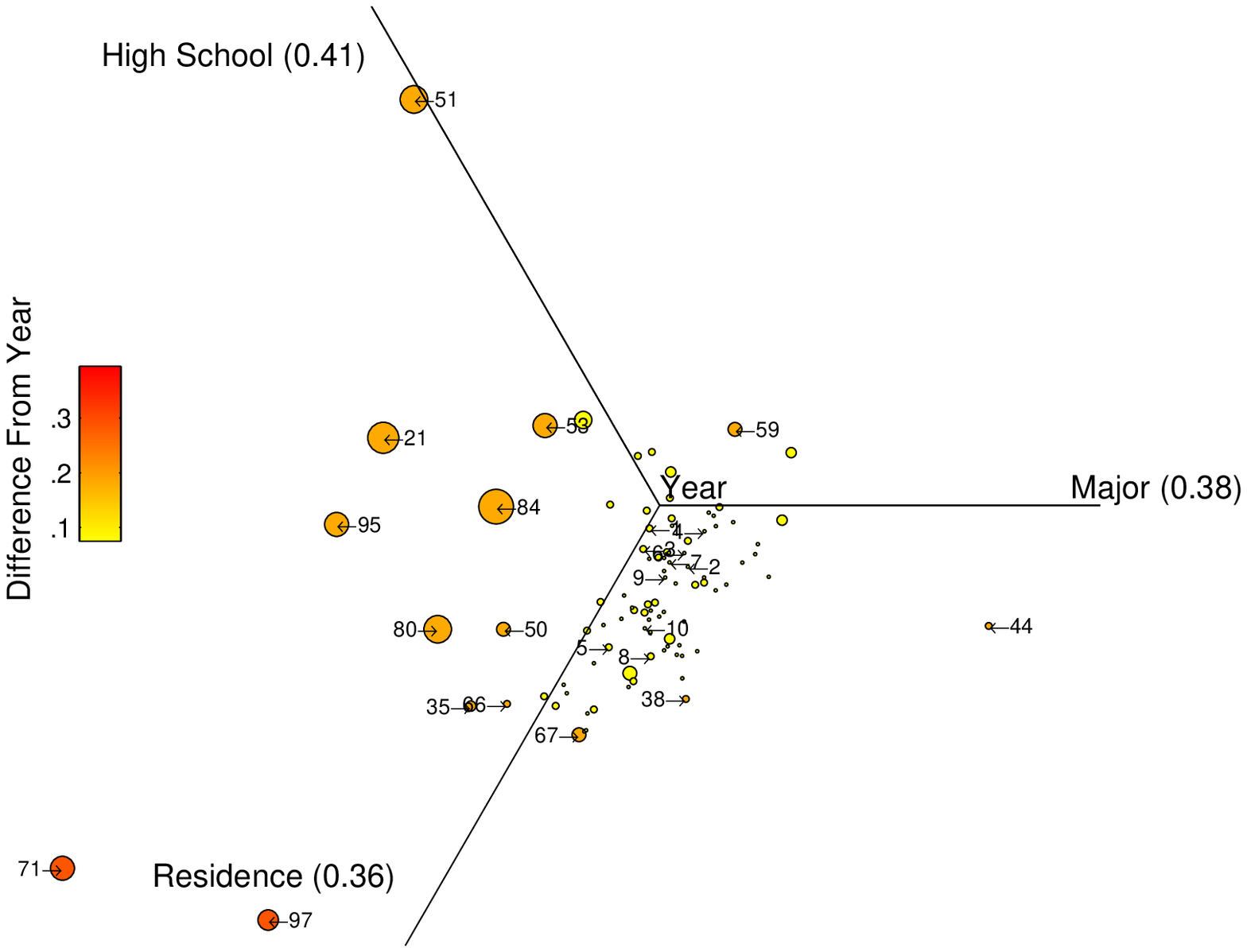}}
\caption{[Color online] (Upper Left) Social organization tetrahedron for the community structures of the Female component of the networks for each of the 100 institutions.  Lighter disks indicate an organization that is based more predominantly on class year.  See the main text for a description of this figure.  (Lower Right) Magnification near the Year vertex.  As in the two previous figures, the disk sizes indicate the maximum distances between partitions.  %In this figure, there are nine cases where $d$ exceeds $.4$.  Four cases have the maximum distance between partitions 1 and 4, UF 21, FSU 53, Texas 84, and Tennessee 95.  Three cases have their maximum distance between partitions 1 and 2, USF 51, UCF 52, and Texas 80.  One case has its maximum distance between partitions 2 and 6, Auburn 71.  The last case has its maximum distance between partitions 1 and 4, Oklahoma 97.
}
\label{fig:Girl}
\end{figure}

\begin{figure}[htbp]
\vspace*{-0.25in}\hspace*{-0.75in}{\includegraphics[width=.85\textwidth]{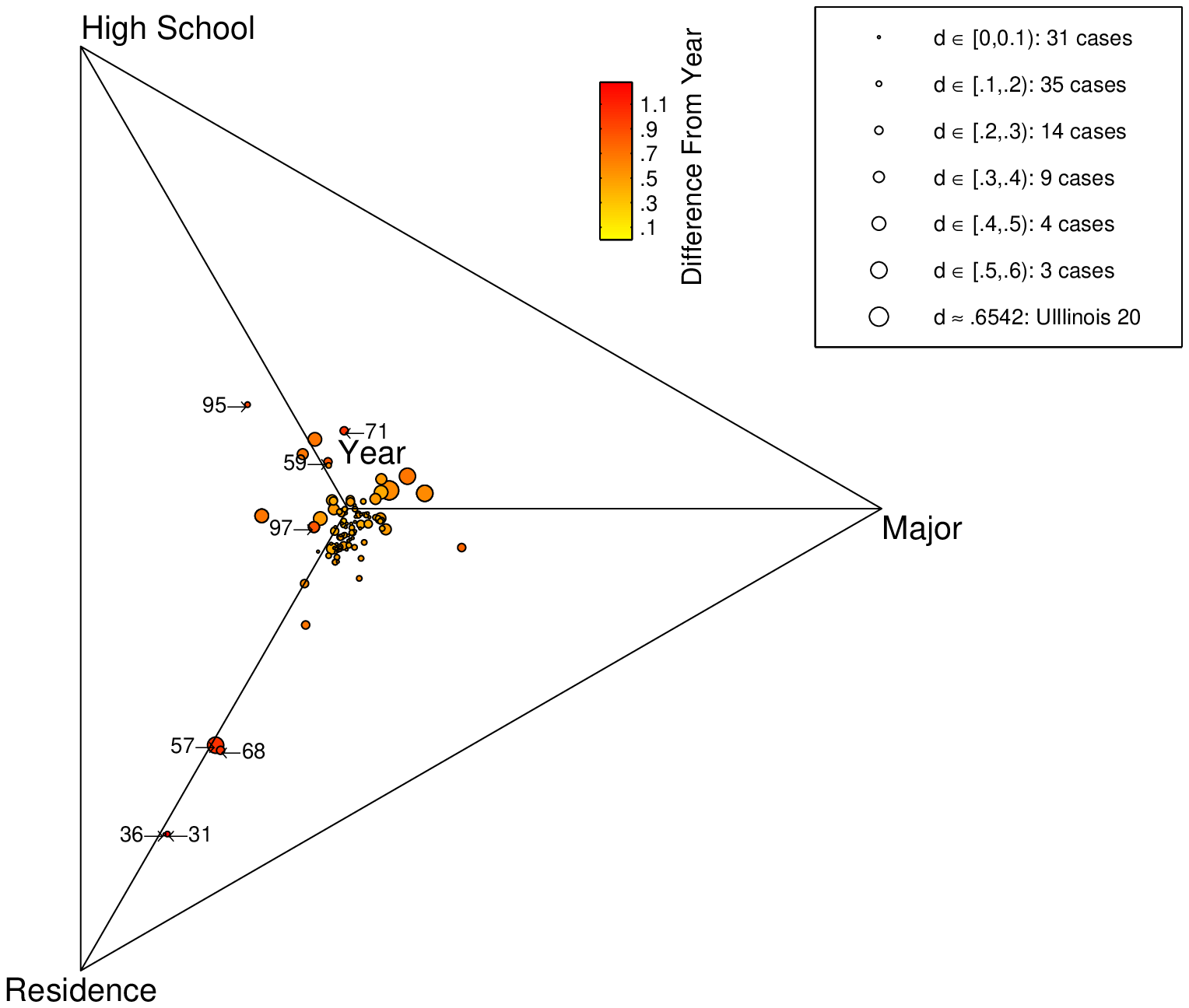}}
\vspace*{-1.05in}\\
\hspace*{1.5in}{\includegraphics[width=.85\textwidth]{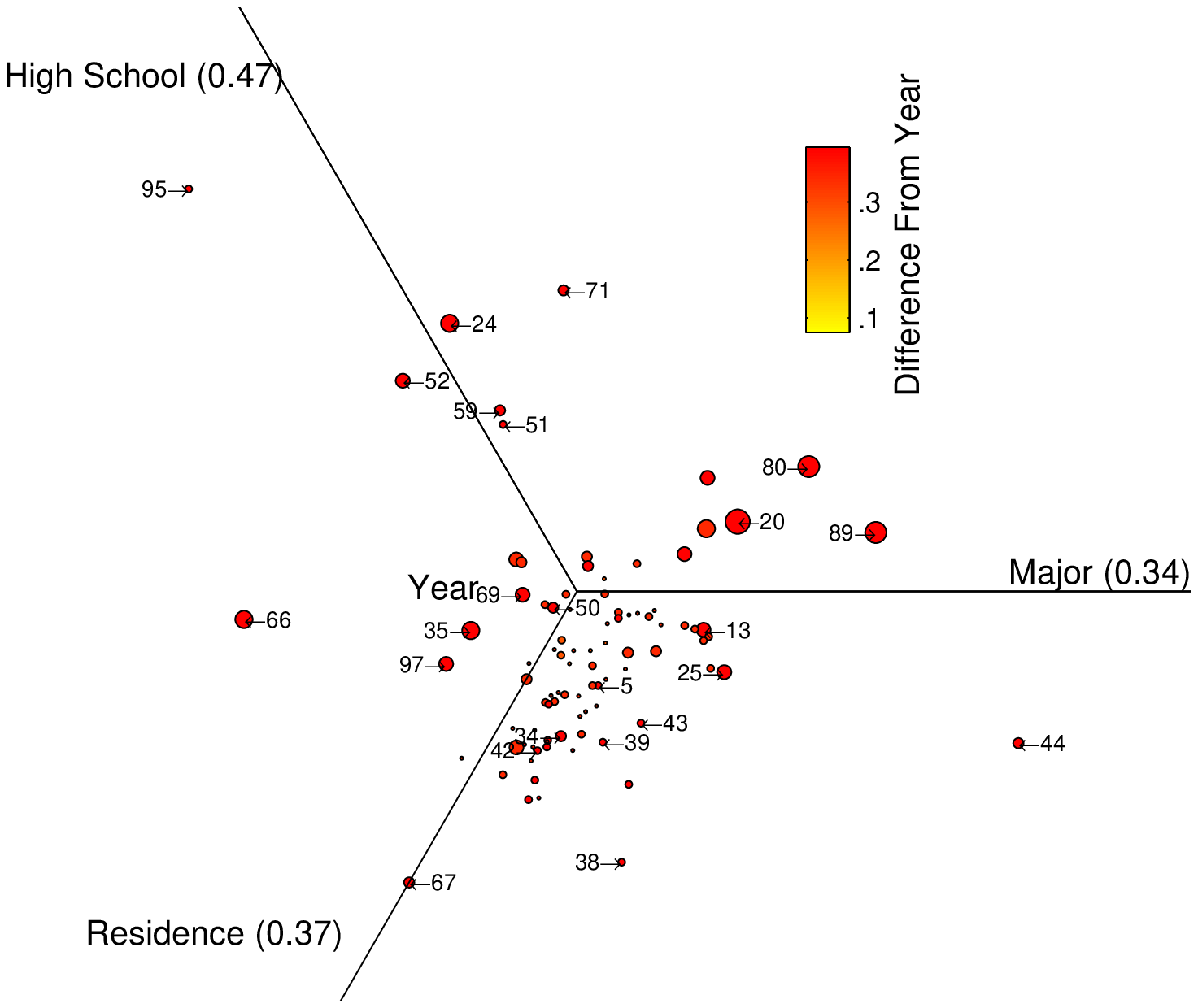}}
\caption{[Color online] (Upper Left) Social organization tetrahedron for the community structures of the Male component of the networks for each of the 100 institutions.  Lighter disks indicate an organization that is based more predominantly on class year.  See the main text for a description of this figure.  (Lower Right) Magnification near the Year vertex.  As in the three previous figures, disk size indicates the maximum distance between partitions. We note that there are more $d>.2$ cases here than in the previous figures, which illustrates the greater variability in the relative positions of the $z$-scores in the different Male networks than was the case for the Full, Student, and Female networks.
%For this figure, there are eight cases in which maximum distance, $d$, exceeds $.4$.  Three cases have the maximum distance between partitions 1 and 6, Virginia 63, Mississippi 66, and Rutgers 89.  Two cases have the maximum distance between partitions 2 and 6, UIllinois 20 and Texas 80.  One case has its maximum distance between partitions 4 and 5, Notre Dame 57.  One case has its maximum distance between partitions 1 and 2, USC 35.  The last case has its maximum distance between partitions 3 and 5, MSU 24.
}
\label{fig:Boy}
\end{figure}

Because of the strong role of class year, we visualize the tetrahedra from a perspective located above the vertex corresponding to class year and project the result into the opposing face of the tetrahedron. We calculate the point $X$ for each of the 6 algorithmic partitions of each network (i.e., using the aforementioned 6 different community detection methods). For each institution, we plot a disk whose center lies at the midpoint of these 6 $X$ coordinates.  The width of each disk is proportional to the maximum observed difference between these 6 sets of coordinates (with these distances separated into bins of width $.1$, as indicated in the legends of Figures \ref{fig:All}--\ref{fig:Boy}).  For example, in Figure \ref{fig:All}, the Pepperdine (86) results have a maximum distance of $.0141$ between partitions, so Pepperdie (86) is represented by one of the smallest disks. Harvard (1) has a maximum distance of .1581 between partitions; this lies in $[.1,.2)$, so Harvard (1) is represented by one of the disks of second smallest size. We emphasize that the computed differences are much larger than the span of the depicted disks, whose sizes allow one to discern the results from different institutions.  
%Only the largest distances indicated in the figure legend have the potential to impact qualitative conclusions about the vertex to which the institution lies closest.

In Figures \ref{fig:All}--\ref{fig:Boy}, we show each of the 100 institutions, identified by number (see Table \ref{tab:CharTable}), using a disk that we have color-coded according to the Cartesian distance of its center from the Year vertex. Class year is the predominant organizing category among the ones present in the data, so most of the institutions are located very close to the Year vertex.  We zoom in on the Year vertex for each figure in order to better discern the relative importance of class year at the institutions.  Importantly, the social organization of a few institutions differs considerably from that of the majority.  Each of these institutions lies close to the Residence
vertex, so their community structures are organized
predominantly according to dormitory residence.  Foremost among these institutions
are Rice (31) and California Institute of Technology (36).  As we
discussed in \cite{siamface}, California Institute of Technology
(Caltech) is well-known to be organized almost exclusively according
to its undergraduate ``House" system \citep{legends}.

In repeatedly observing a strong correlation of class year with community structure, it is relevant to recall that the community detection method that we have employed optimizes modularity at the default resolution. Because of the resolution limit of modularity \citep{resolution}, it might be interesting to explore individual networks at different scales using resolution parameters \citep{spinglass,santobig,ourreview}.   We reiterate, however, that our focus in the present paper is on large-scale features rather than precise node membership of network partitions.

In Figure \ref{fig:All}, we show the social organization tetrahedron for
the Full networks (i.e, for the the largest connected components of
the complete networks) for each institution.  Although
the community structure of nearly all of the Full networks are
organized overwhelmingly by class year, a few of them are also heavily
influenced by dormitory residence.  (We already mentioned above that
Rice (31) and Caltech (36) are organized predominantly by Residence.)
For example, dormitory residence also
dominates the community structure at UC Santa Cruz [UCSC] (68), though to a lesser extent than at Rice and Caltech.  We also observe relatively high Residence $z$-scores at Smith (60), Auburn (71), and 
University of Oklahoma (97).  Major seems to be most important relative to the other available characteristics at
Oberlin (44) and Maine (59), though in both cases its relative correlation 
pales in comparison to that of class year.  High School
seems to be most important at USF (51) and Tennessee (95), though
class year is again more important.  Most of the institutions are
clustered tightly near the Year vertex, but Residence can often be
rather important (and sometimes even the most important category, as
we have seen in three cases).

In Figure \ref{fig:Student}, we show the social organization tetrahedron for the Student networks (i.e., for the largest connected component of the student-only subnetworks) for each institution.  As we saw with the Full networks, most of the institutions have community structures that are organized overwhelming according to class year.  Rice, Caltech, Smith, UCSC, Auburn, and Oklahoma are again exceptions, as dormitory residence also exerts considerable (or even primary) influence at these institutions.  Additionally, considering the Student network reduces the relative dominance of the Year vertex, although it clearly still dominates the social organization.  This feature is illustrated by institutions such as UC (64), UF (21), and Rutgers (89).  

In Figure \ref{fig:Girl}, we show the social organization tetrahedron for the Female networks (i.e., for the largest connected component of the female-only subnetworks) for each institution.  Class year is once again the overwhelmingly dominant organizing characteristic, and dormitory residence is again important at institutions such as Rice, Caltech, Smith, UCSC, Auburn, and Oklahoma.  However, we now observe an increased importance of the High School vertex.  USF (51), Tennessee (95), UF (21), FSU (53), and GWU (54) all lie closer to the High School vertex than was the case in the Full and Student networks.

In Figure \ref{fig:Boy}, we show the social organization tetrahedron for
the Male networks (i.e., for the largest connected component of the
male-only subnetworks) for each institution.  Class
year is once again the overwhelmingly dominant organizing
characteristic, and dormitory residence is again the most important
category at institutions such as Rice, Caltech, and UCSC.
Interestingly, considering the Male network suggests that Residence is
the most important factor for the social organization for the males at
Notre Dame (57).  Residence also exerts an important influence on
the males at Michigan (67).  This is starkly different from what we
observed for these institutions in the Full, Student, and Female
networks (and would seem to be something
interesting to investigate more thoroughly in the future using other
data and methods).
The Male UCF (52), MSU (24), USF (51), Auburn (71), and
Maine (59) networks are strongly influenced by High School.  The Male
networks at Texas (80), Rutgers (89), and University of Illinois at
Urbana-Champaign (20) stand out from other universities because of
their proximity to the Major vertex.

%%%%%%

\subsection{Discussion}

As described above, we see using the $z$-scores of the Rand coefficients for demographic characteristics versus algorithmic community assignments that Year is the strongest organizing factor at most institutions but that Residence is much more important for the community organization at some institutions than at others.  The correlation with Residence is especially prominent at Rice (31) and Caltech (36).  We also observe that the Male networks tend to be more scattered around Year, as some institutions exhibit a stronger correlation with Major, whereas others have a stronger correlation with high school.  This suggests that there are potential differences in the gender patterns of friendships, which would be interesting to investigate in future studies with new data.  We do not explore this general issue further and instead attempt to identify interesting comparisons with the results that we obtained above.  Although it is of course impossible to be exhaustive in our observations, we present all of our assortativity values, regression model coefficients, and community-comparing $z$-scores in the tables in Appendix A. We also highlight some interesting facets of our results.

Of particular interest is the comparison of results from the dyad-level regression models to those from community-level correlations. We note in particular that the logistic regression and exponential random graph models that we employed for the smallest 16 institutions specify that almost all institutions and all of their subnetworks give the highest model coefficient contribution towards a link between nodes from a common High School. However, as we have seen---and which is particularly evident using the visualizations with tetrahedra---at the community level, most institutions are organized by class year and have a  relatively small correlation with high school.

Even in the rare cases in which the rank ordering of the four
correlations (with Year, Residence, Major, and High School) at the community level matches that obtained via dyad-level model coefficients, such as with the logistic regression
model for the Full and Female networks from Caltech (36), the relative
sizes of the contributions at the dyad level are completely
different from those observed at the community level. 
Caltech supplies an illustrative example of the different insights obtained from community-detection versus logistic regression and exponential random graph models both because of its small size and because of its outlying 
%outlier  {\bf map: are we allowed to use outlier as an adjective?}
correlation with
dormitory residence at the community level.  A simple
interpretation of the apparent dichotomy between the dyad-level model
coefficients and the correlations at the community scale is that the
presence of two students from the same high school at a small institution like Caltech yields a significant increase in the likelihood of a tie between those students. Even though the corresponding model coefficient is smaller than in any of the other of the 16 smallest institutions, it is comparable to that for common residence (called ``Houses'' at Caltech). Nevertheless,
the very small number of node pairs at Caltech that have the same high school relative to the total
number of node pairs has a very small effect at the community
level, as the algorithmically obtained communities are correlated overwhelmingly with House affiliation. The ERGM result with triangle contributions
makes this distinction even more striking, as the common high
school coefficient is actually larger than the coefficient from common
House.

We also observe other features that might be worthy of future investigation using other data sets and methodologies.   We report the results of our calculations in depth in Tables \ref{tab:CharTable}--\ref{tab:ZTable}.  Here we highlight only a few potentially interesting examples in which different methods or different subnetworks yield apparently different qualitative conclusions.  For example, we found that Major is the second most important factor for the organization of the communities in all of the Oberlin (44) networks, but only for the Full and Male networks does the logistic regression give the second highest coefficient for Major.  We also observed that the relative ordering of Major at the same institution is sometimes gender-dependent.  For example, Major gives the second largest $z$-score in the Female and Male networks of Stanford (3), but it gives the fourth largest $z$-score in Stanford's Full network.  Even more interesting, Major gives the second largest $z$-score for the Female network at UVA (16), the third largest $z$-score for UVA's Male network, and the fourth largest $z$-score for its Full network.
%UVA (16), UChicago (30), and UConn (91), 
%and in the Male networks of NYU (9), UCSB (37), Carnegie (49), and Rutgers (89). 
The communities in the Auburn (71) Female
network are dominated by Residence, but those in the other Auburn networks are not.  Similarly, the communities in the MIT (8) Male network are dominated by Residence, but those in the other MIT (8) networks are not.  Another interesting disparity based on gender occurs in the communities in the Tennessee (95) Full and Student networks, which have their second largest contributions from High School, whereas those in the other two Tennessee networks have their second largest contributions from Residence. 
%(recalling that the Male and Female subnetworks include only same-gender links).

%%%%%%%%%%%

\section{Conclusions}

We have studied the social structure of Facebook ``friendship''  networks at one hundred American institutions at a single point in time (using data from September 2005).  To compare the organizations of the 100 institutions using categorical data, we considered both microscopic and macroscopic perspectives.  In particular, calculating assortativity coefficients and regression model coefficients based on observed ties allows one to examine homophily at the local level, and algorithmic community detection allows a complementary macroscopic picture.
These approaches complement each other, providing different perspectives on investigations of these Facebook networks. Such complementary calculations are particularly valuable when the microscopic and macroscopic perspectives identify different dominant contributions.  For example, in the Caltech networks, the assumed ground truth of the importance of the House system is captured better by computing community structure.

This ``real-world ensemble'' of 100 networks formed by ostensibly similar 
%(but not identical) 
mechanisms has the potential to provide a testing ground for various models of network formation.  Because of the useful comparisons such an ensemble of data can facilitate, this data will similarly be useful for studies of dynamic processes on networks, algorithmic community detection, and so on.  Because of the different rates of initial Facebook adoption at different institutions, the single point in time represented by the data might usefully describe different stages in the formation of an online social network. In order to pursue such ideas further, one needs to start by studying the networks for their own sake and comparing their structures.  This was the goal of the present paper. In particular, we have identified some of the key differences across these 100 realizations of online social networks.

Some of our observations confirm conventional wisdom or are intuitively clear, providing soft verification of our investigation via expected results.  For example, we found that class year is often important, Houses are important at Caltech, and high school plays a greater role in the social organization of large universities than it does at smaller institutions (where there are typically fewer pairs of people from the same high school).  Other results are quite fascinating and merit further investigation.  In particular, the differences in the community structures of the female-only and male-only networks would be interesting to investigate in both offline and online settings. The Facebook data suggests that women are typically more likely to have friends within their common residence (among the demographic data to which we have access) but that the characteristics in the communities in the male-only networks exhibit a wider variation. Investigating this thoroughly would require different data sets and methodologies, especially if one wishes to discern the causes of such friendships from observed correlations. 

The Facebook networks that we study offer imperfect representations of corresponding real-life social networks, which have different properties from online social networks.  It is thus crucial that our results are complemented by studies of the corresponding real networks in order to quantify the extent of such differences.

%%%%%%%%%%

\section*{Acknowledgements}

We thank Adam D'Angelo and Facebook for providing the data used in this study.  We also acknowledge Sandra Gonz\'{a}lez-Bail\'{o}n and Erik Kelsic for useful discussions.  We thank Christina Frost for developing some of the graph visualization code that we used (available at {\tt http://netwiki.amath.unc.edu/VisComms}).  ALT was funded by the NSF through the UNC AGEP (NSF HRD-0450099) and by the UNC ECHO program.  PJM was funded by the NSF (DMS-0645369) and the UNC ECHO program.  MAP acknowledges a research award (\#220020177) from the James S. McDonnell Foundation.

%%%%%%%%

\clearpage{}

%%%%%%%%

\clearpage
%Moved Appendix to new document
\appendix

\section{Tables}

In Table \ref{tab:CharTable}, we give for each of the 100 institutions the numbers of nodes and edges for each of the Facebook networks (and subsets thereof) that we have investigated.  
%In Table \ref{tab:AssortTable}, we give the assortativity values by Major, Residence, Year, and High School for each of these networks and also by Gender for the Full and Student networks.
In Table \ref{tab:AssortTable}, we give the assortativity values for each of the networks.  For each institution, we calculate assortativity values for Gender only for the Full and Student network subsets.  We calculate Major, Residence, Year, and High School assortativity values for each of the four network subsets (Full, Student, Female, and Male). 

Recall that we studied regression models for the 16 institutions with the smallest Facebook networks.  In Table \ref{tab:LogOdds1}, we report the results of a logistic regression model with {\tt edge} and {\tt nodematch} terms.  (All coefficients differ from zero with $p$-values less than $1 \times 10^{-4}$.)  In Table \ref{tab:ZTable}, we similarly report the results of an ERGM that supplements the logistic regression model with {\tt triangle} terms.  (Again, all resulting model coefficients differ from zero with a $p$-value of less than $1 \times 10^{-4}$.)

In Table \ref{tab:ZTable}, we report the maximum $z$-score for each demographic category that we obtained from the 6 different community detection partitions (described in the text) of each Facebook network (and their subsets) compared to categorical partitions based on each of Major, Residence, Year, and High School.  We divide the networks in this table into five sections: (1) networks for which the High School category gives the highest $z$-score; (2) networks for which the Residence category gives the highest $z$-score; (3) networks for which Year gives the highest $z$-score and High School gives the second highest; (4) networks for which Year gives the highest $z$-score and Major gives the second highest; and (5) networks for which Year gives the highest $z$-score and Residence gives the second highest.

{\footnotesize

\begin{landscape}
	\centering
\begin{longtable}{|l|l|l|l|}
			\caption{Characteristics for each of the networks and subnetworks: institution name, the identifying number given by Facebook, number of nodes in each network and subnetwork, and the number of edges in each network and subnetwork.} \label{tab:CharTable}\\
			\hline
		 \multicolumn{1}{|c|}{\textbf{Institution}} & \multicolumn{1}{|c|}{\textbf{Number}} & \multicolumn{1}{|c|}{\textbf{Nodes (Full, Student, Female, Male)}} & \multicolumn{1}{|c|}{\textbf{Edges (Full, Student, Female, Male)}} \\ \hline 
\endfirsthead

\multicolumn{4}{c}%
{{\bfseries \tablename\ \thetable{} -- Continued from previous page}} \\[0.5ex]
 \\ \hline
 \multicolumn{1}{|c|}{\textbf{Institution}} & \multicolumn{1}{|c|}{\textbf{Number}} & \multicolumn{1}{|c|}{\textbf{Nodes (Full, Student, Female, Male)}} & \multicolumn{1}{|c|}{\textbf{Edges (Full, Student, Female, Male)}} \\ \hline

\endhead

%This is the footer for all pages except the last page of the table...
\\ \hline
  \multicolumn{4}{|l|}{{Continued on Next Page\ldots}} \\ \hline
\endfoot

%This is the footer for the last page of the table...
\\ \hline
  \multicolumn{4}{|l|}{{}} \\ \hline
\endlastfoot

		Harvard & $1$ & $(15086,7425,5865,6850)$ & $(824595,404415,173639,187742)$ \\ \hline
		Columbia & $2$ & $(11706,8057,5864,4209)$ & $(444295,296971,135234,76037)$ \\ \hline
		Stanford & $3$ & $(11586,7183,4562,5501)$ & $(568309,345561,132904,135932)$ \\ \hline
		Yale & $4$ & $(8561,5405,3572,3891)$ & $(405440,258886,85133,95992)$ \\ \hline
		Cornell & $5$ & $(18621,12843,8028,8538)$ & $(790753,511386,203303,171118)$ \\ \hline
		Dartmouth & $6$ & $(7677,4705,3052,3417)$ & $(304065,176665,68675,70858)$ \\ \hline
		UPenn & $7$ & $(14888,10106,6405,6625)$ & $(686485,446037,172277,150449)$ \\ \hline
		MIT & $8$ & $(6402,4283,2298,3359)$ & $(251230,158838,58906,70094)$ \\
\end{longtable}
\end{landscape}

\begin{center}
	\centering
		\begin{longtable}{|l|l|l|l|l|}
\caption{Assortativity values for each category for each of the 4 networks (Full, Student, Female, and Male) for each of the 100 institutions.   We only calculate assortativity by Gender for the Full and Student networks.  (We leave blank spots in the corresponding table entries for the Male and Female networks.)}
\label{tab:AssortTable}\\	
\hline
		 \multicolumn{1}{|c|}{\textbf{Institution No.}} & \multicolumn{1}{|c|}{\textbf{Full}} & \multicolumn{1}{|c|}{\textbf{Student}} & \multicolumn{1}{|c|}{\textbf{Female}} & \multicolumn{1}{|c|}{\textbf{Male}} \\ \hline 
\endfirsthead

\multicolumn{5}{c}%
{{\bfseries \tablename\ \thetable{} -- Continued from previous page}} \\[0.5ex]
 \hline
 \multicolumn{1}{|c|}{\textbf{Institution}} & \multicolumn{1}{|c|}{\textbf{Full}} & \multicolumn{1}{|c|}{\textbf{Student}} & \multicolumn{1}{|c|}{\textbf{Female}} & \multicolumn{1}{|c|}{\textbf{Male}} \\ \hline

\endhead

%This is the footer for all pages except the last page of the table...
\hline
  \multicolumn{5}{|l|}{{Continued on Next Page\ldots}} \\ \hline
\endfoot

%This is the footer for the last page of the table...
\hline
  \multicolumn{5}{|l|}{{}} \\ \hline
\endlastfoot

Harvard $1$ & & &   \\ \hline
Gender & $0.058144$ & $0.049178$ & & \\
Major & $0.056293$ & $0.046659$ & $0.051852$ & $0.064064$ \\
Residence & $0.14679$ & $0.11951$ & $0.13803$ & $0.15431$ \\
Year & $0.47981$ & $0.60723$ & $0.4871$ & $0.44035$ \\
High School & $0.023132$ & $0.02419$ & $0.024247$ & $0.026473$ \\ \hline \hline
Columbia $2$ & & &   \\ \hline
Gender & $0.087283$ & $0.085847$ & & \\
\end{longtable}
%\label{tab:AssortTable}
%\caption{Assortativity values for each category for each of the 4 networks (Full, Student, Female, and Male) for each of the 100 institutions.   We only calculate assortativity by gender for the Full and Student networks. \todo{Mandi, this caption always gives me errors and is placed very badly --- please fix}}
\end{center}

{%\scriptsize
%I have the other data, it is in the excel file, I will put this data in when we decide on how we want to change this table
\begin{landscape}
		\begin{longtable}{|c|c|c|c|c|c|}
%		\centering{\bfseries \tablename\ \thetable{} }
		\caption{Logistic regression coefficients for a model combining a density ({\tt edge}) term and {\tt nodematch} contributions for the increased propensity of two nodes with the same categorical value to have an edge connected between them.  This is calculated individually for Year, Residence, High School, and Major.  We give the standard error for each coefficient in parentheses. All coefficients are statistically significantly different from zero with $p$-value less than $1 \times 10^{-4}$. Wellesley (22), Smith (60), and Simmons (81) are female-only institutions, so we list the values for their Male networks as NA.    
} 	\label{tab:LogOdds2} \\
			\hline
Wellesley $22$ & Edge & Year & Residence & High School & Major   \\ \hline
Full & $-4.4291(0.0047086)$ & $1.8249(0.0067097)$ & $1.2546(0.012002)$ & $3.1738(0.041398)$ & $0.70232(0.013501)$ \\
Student & $-4.3656(0.0063232)$ & $1.7437(0.0082165)$ & $1.2512(0.013129)$ & $3.2966(0.05217)$ & $0.62071(0.01745)$ \\
Female & $-4.4673(0.0054048)$ & $1.8587(0.0073823)$ & $1.2749(0.012577)$ & $3.19(0.044896)$ & $0.66471(0.014677)$ \\
Male & NA & NA & NA & NA & NA \\ \hline
Caltech $36$ & Edge & Year & Residence & High School & Major   \\ \hline
Full & $-3.6903(0.012891)$ & $1.5382(0.018233)$ & $2.4151(0.018644)$ & $2.3789(0.14869)$ & $0.53388(0.02881)$ \\
\hline
\end{longtable}
	
\end{landscape}  
}

{\scriptsize
\begin{landscape}
		\begin{longtable}{|c|c|c|c|c|c|c|}
%		\centering{\bfseries \tablename\ \thetable{} }
		\caption{ERGM coefficients for the model (described in the text) that combines density ({\tt edge}) and {\tt triangle} terms with {\tt nodematch} contributions representing the increased propensity for two nodes with the same categorical value to have an edge connected between them.  (This is calculated individually for Year, Residence, High School, and Major.) We give the standard error for each coefficient in parentheses. All coefficients are statistically significantly different from zero with $p$-value less than $1 \times 10^{-4}$. Wellesley (22), Smith (60), and Simmons (81) are female-only institutions, so we list the values for their Male networks as NA.    
} 	\label{tab:LogOdds1} \\
			\hline
Wellesley $22$ & Edges  & Triangles & Year & Residence & High School & Major   \\ \hline
Full & $-5.5166(0.29946)$ & $0.18714(0.00040795)$ & $1.0432(1.1574)$ & $1.2079(0.014731)$ & $3.612(8.5012)$ & $0.58573(0.01648)$ \\
Student & $-5.395(0.40299)$ & $0.18873(0.00054665)$ & $0.89815(0.93197)$ & $1.262(0.83757)$ & $3.7139(8.9764)$ & $0.44432(1.2761)$ \\
Female & $-5.5395(0.47528)$ & $0.20854(0.00050963)$ & $1.0713(0.70673)$ & $1.2339(0.53312)$ & $3.6698(6.0111)$ & $0.61932(1.858)$ \\
Male & NA & NA & NA & NA & NA & NA \\ \hline
Caltech $36$ & Edges  & Triangles & Year & Residence & High School & Major   \\ \hline
Full & $-4.9776(0.0013776)$ & $0.17766(1.64e-005)$ & $0.99434(0.0014976)$ & $1.1638(0.0010284)$ & $2.8536(0.087757)$ & $0.64673(0.0021013)$ \\
\hline
\end{longtable}	
\end{landscape}

}

%BEGIN TABLE

\begin{center}
	\centering
		\begin{longtable}{|l|l|l|l|l|}
\caption{Maximum $z$-scores of the Rand coefficient obtained for the 6 employed community detection algorithms (see the discussion in the text) for each categorical variable in every network (Full, Student, Female, and Male) for each of the 100 institutions.  We italicize $z$-scores that are less than 2.  We divide the table into five parts: (1) networks in which High School yields the highest $z$-score, (2) networks in which Residence yields the highest $z$-score, (3) networks in which Year yields the highest $z$-score and High School yields the second highest $z$-score, (4) networks in which Year yields the highest $z$-score and Major yields the second highest $z$-score, and (5) networks in which Year yields the highest $z$-score and Residence yields the second highest $z$-score.}\label{tab:ZTable}\\

		\hline
		 \multicolumn{1}{|c|}{\textbf{Institution and Network}} & \multicolumn{1}{|c|}{\textbf{Major}} & \multicolumn{1}{|c|}{\textbf{Residence}} & \multicolumn{1}{|c|}{\textbf{Graduation Year}} & \multicolumn{1}{|c|}{\textbf{High School}}\\ \hline 
\endfirsthead

\multicolumn{5}{c}%
{{\bfseries \tablename\ \thetable{} -- Continued from previous page}} \\[0.5ex]
   \hline
 \multicolumn{1}{|c|}{\textbf{Institution and Network}} & \multicolumn{1}{|c|}{\textbf{Major}} & \multicolumn{1}{|c|}{\textbf{Residence}} & \multicolumn{1}{|c|}{\textbf{Graduation Year}} & \multicolumn{1}{|c|}{\textbf{High School}} \\ \hline

\endhead

%This is the footer for all pages except the last page of the table...
 \hline
  \multicolumn{5}{|l|}{{Continued on Next Page\ldots}} \\ \hline
\endfoot

%This is the footer for the last page of the table...
 \hline
  \multicolumn{5}{|l|}{{}} \\ \hline
\endlastfoot
\hline
\hline
\emph{High School:} &  & & & \\ \hline
\textbf{Auburn 71 Male} & $37.6893$ & $14.8497$ & $42.7784$ & $70.6776$ \\ \hline
\textbf{Tennessee 95 Male} & $15.6741$ & $20.4034$ & $42.8019$ & $58.5508$ \\ \hline
\hline
\emph{Residence:} &  &  & & \\ \hline
\textbf{Rice 31 Full} & $15.3123$ & $1404.4502$ & $196.14$ & $4.3858$\\ \hline
\textbf{Caltech 36 Full} & $4.0649$ & $222.9566$ & $8.5967$ & $4.9078$\\ \hline
\textbf{UCSC 68 Full} & $36.7889$ & $945.8502$ & $481.3222$ & $10.3043$\\ \hline
\textbf{Rice 31 Student} & $19.2677$ & $1523.4423$ & $137.8101$ & $\mathit{1.6253}$ \\
\hline
\end{longtable}
\end{center}

}

%\bibliographystyle{model2-names}
%\bibliography{Mastersbib}

\end{document}